\documentclass[twocolumn,prd,floatfix]{revtex4}
\usepackage{physics}
\usepackage{color}
\usepackage{tabularx}
\usepackage{epsfig}
\usepackage{amsmath}
\usepackage{amssymb}
\usepackage{bm}
\usepackage{graphicx}
\usepackage{multirow}
\usepackage{mathtools} 
\usepackage{xcolor} 

\begin{document}
	
	\title{Improving Gravitationally Induced Entanglement in Interferometric Configurations}
	
	\author{Alireza Maleki}
	\affiliation{School of Physics, Institute for Research in Fundamental Sciences (IPM), Tehran, Iran}
\begin{abstract}
	Proposals to detect gravitationally induced entanglement between massive particles offer a promising pathway toward probing the quantum nature of gravity. We perform a systematic geometric optimization and comparative analysis of Stern--Gerlach (SG) and Mach--Zehnder (MZ) matter-wave interferometers for this purpose, deriving exact analytical expressions for the gravitational phase of each configuration. Our first key result is that geometric optimization is indispensable. For commonly assumed experimental parameters, neglecting the optimization of geometric ratios can suppress the accumulated phase, and hence the detectability of entanglement, by up to several orders of magnitude. Second, we provide a complete comparative analysis of two SG orientations (linear and parallel) and three MZ orientations (translation-symmetric, mirror-symmetric, and parallel-arms). The parallel-arms SG and MZ configurations tend to yield the largest phase accumulations and are nearly equivalent in performance, whereas the translation-symmetric MZ setup exhibits destructive-interference nulls that act as blind operating points where detection becomes ineffective. These observations highlight the importance of geometry-aware design. A Monte Carlo error analysis of realistic fabrication tolerances indicates that the parallel-arms configurations are generally more robust, maintaining low relative phase uncertainty over broad parameter intervals. Furthermore, an evaluation of Bell violations assesses the feasibility and signal-to-noise ratio of each configuration. Our results identify geometric optimization, particularly through parallel-arms designs, as an important consideration for near-term experiments targeting gravitationally induced entanglement.
\end{abstract}
	\maketitle
	
	\section{INTRODUCTION}
	\label{sec:introduction}
	
	For over a century, quantum mechanics and general relativity have stood as the twin pillars of modern physics, together describing phenomena from subatomic to cosmic scales \cite{mukhanov2007introduction,zee2013einstein,buchbinder2021introduction,maleki2020speed,maleki2024cosmic,maleki2026area}. Yet these frameworks remain fundamentally incompatible. At the Planck scale, general relativity loses predictive power while quantum theory provides no description of spacetime geometry, giving rise to conceptual puzzles such as the black hole information paradox \cite{hawking1974black,susskind2005introduction,marletto2025quantum}.
	
	Despite numerous theoretical efforts to quantize gravity, experimental verification remains unattainable due to the inaccessibility of Planck-scale energies \cite{alfaro2005quantum}. This empirical barrier has led some physicists, notably Freeman Dyson, to question whether the quantum nature of gravity, like the aether before it, should be abandoned if it lacks experimentally accessible meaning. In response, alternative frameworks have emerged in which gravity is an emergent phenomenon rather than a fundamental interaction \cite{padmanabhan2015emergent,verlinde2017emergent}. These developments underscore the urgent need for experiments that can distinguish between classical and quantum descriptions of gravity.
	
	Quantum information science has opened new experimental avenues, with quantum interferometry at the forefront \cite{peres2004quantum,lunghi2013experimental}. Significant experimental progress has already been made in investigating gravitational effects, ranging from precision tests of gravity to gravitational wave detection using various atom interferometry methods \cite{zych2011quantum, tino2021testing, stray2022quantum,schnabel2010quantum}. More recently, quantum metrology and interferometry are increasingly being directed toward fundamental questions, specifically the question of whether gravity itself obeys the laws of quantum mechanics \cite{bose2025massive, belenchia2016testing, biswas2023gravitational, miao2020quantum, haine2021searching, wakakuwa2026detectability, agafonova2026one, an2026test}.
	
	Proposals exploiting gravitational phase shifts between spatially superposed massive particles in the \(10^{-14}\) to \(10^{-12}\) kg range aim to detect entanglement generated solely by gravitational interaction \cite{bose2017spin,marletto2017gravitationally}. The observation of such entanglement would constitute a definitive signature of non-classical gravity \cite{bose2025massive,marletto2025classical,krisnanda2020observable,howl2021non,torovs2021relative,carney2021using,maleki2022complementarity,weber2024bose,hanif2024testing,noman2026operational,ming2026unveiling}. However, the experimental realization of these schemes faces severe challenges \cite{arndt2014testing,pedernales2022enhancing,pedalino2026probing}. The weak gravitational signal can be masked by Casimir--Polder forces, electrostatic and magnetic dipolar couplings, and environmental noise \cite{gunnink2023gravitational,schut2023relaxation,bose2025massive,chen2025gravitationally}. Mitigating these effects requires ultra-high vacuum, cryogenic temperatures, and advanced vibration isolation \cite{bose2025massive,marletto2025quantum,raman2025massive,gerlich2011quantum,pedalino2023experimental}. Despite these hurdles, interferometry with large quantum systems remains a rapidly developing field \cite{pedalino2026probing,wan2016free,fein2019quantum} and has recently demonstrated spatial superpositions and matter-wave interference for molecules with masses up to \(10^{-21}\,\mathrm{kg}\) \cite{pedalino2026probing,margalit2021realization}, marking steady progress toward the required target regime.

	Two distinct interferometer geometries have dominated recent proposals: the Stern--Gerlach (SG) and Mach--Zehnder (MZ) interferometers. The two are suited to different classes of quantum systems. The SG geometry relies on a spin degree of freedom to split and recombine the wavepacket coherently, making it naturally compatible with spin-bearing systems such as nitrogen-vacancy centers in nanodiamonds \cite{bose2017spin}. The MZ geometry splits the wavepacket using beam splitters rather than spin-dependent forces. This design is aimed at accommodating massive candidates independent of their spin, including large molecules and nanoparticles \cite{marletto2017gravitationally}. A point largely absent from existing analyses is the identification of the geometric configuration that maximizes the gravitationally induced phase and hence the detectability of entanglement. \\
Our previous analysis \cite{maleki2026role,maleki2026optimal} showed that geometric optimization is essential in these experiments. Widely used approximations can yield unreliable results or poorly optimized configurations, and specific choices of interferometer parameters can completely eliminate the expected entanglement in certain setups. Here, we extend that work to a systematic comparative study of SG and MZ interferometers, specifically focusing on parallel-arm configurations whose optimization has not yet been studied. We derive exact phase expressions for all five geometries, optimize each over its geometric ratios, and perform a Monte Carlo error analysis to quantify the impact of geometric uncertainties on the accumulated phase. Importantly, we evaluate all configurations at realistic parameter scales commonly considered for such setups. This reveals that suboptimal geometric choices can suppress the detectability of entanglement by up to several orders of magnitude, whereas operating at the optimal geometry substantially enhances the signal. We further find that the parallel-arms configurations generally provide both a larger accumulated phase and a relative phase uncertainty that is comparable to or lower than that of their linear counterparts across most of the accessible parameter space. Despite this shared advantage in phase accumulation, the two optimal designs impose significantly different geometric requirements. The parallel SG configuration necessitates a highly elongated vertical geometry to maximize its projection along the separation axis, whereas the parallel-arm MZ configuration operates optimally as a symmetric square-like interferometer with comparable horizontal and vertical arms.

	\section{COMMON PHASE FORMALISM AND ENTANGLEMENT MEASURES}
	\label{sec:formalism}
	
	We consider two massive particles of masses \(m_1\) and \(m_2\), each traversing its own interferometer (see, e.g., Fig.~\ref{fig:sg-linear}). The two interferometers are assumed to be geometrically identical in shape and size. The upper and lower paths in the \(i\)-th interferometer (\(i=1,2\)) correspond to the orthogonal states \(\vert u_i\rangle\) and \(\vert d_i\rangle\), with reflectivity and transmissivity coefficients \((r_i,t_i)\). After the first beam splitter, the system state is
	\begin{equation}
		\vert \psi(0)\rangle =
		\big(r_1\vert u_1\rangle + t_1\vert d_1\rangle\big)
		\otimes
		\big(r_2\vert u_2\rangle + t_2\vert d_2\rangle\big).
		\label{eq:initial-state}
	\end{equation}
	Gravitational interactions induce a phase shift, as first demonstrated in the Colella--Overhauser--Werner (COW) experiment \cite{colella1975observation}. The accumulated phase is
	\begin{equation}
		\phi_G = -\int \frac{m\Phi}{\hbar}\,dt,
		\label{COWph}
	\end{equation}
	where \(m\) is the particle mass, \(\Phi\) is the gravitational potential, and \(\hbar\) is the reduced Planck constant. Before recombination, the state evolves to
	\begin{align}
		\nonumber
		|\psi(T)\rangle
		&= t_1t_2 e^{i\Phi_{dd}}|d_1\rangle |d_2\rangle
		+ t_1r_2 e^{i\Phi_{du}}|d_1\rangle |u_2\rangle \\
		&\quad
		+ r_1t_2 e^{i\Phi_{ud}}|u_1\rangle |d_2\rangle
		+ r_1r_2 e^{i\Phi_{uu}}|u_1\rangle |u_2\rangle,
		\label{eq:evolved-state}
	\end{align}
	where \(\Phi_{xy}\) denotes the total phase accumulated when particle~1 follows path \(x \in \{u,d\}\) and particle~2 follows path \(y \in \{u,d\}\). These four phases depend on the specific interferometer geometry and are computed in Secs.~\ref{sec:sg} and~\ref{sec:mz} for each configuration.
	
	The nonlocal character of entangled quantum systems was brought to prominence in 1935 by Einstein, Podolsky, and Rosen \cite{einstein1935can}, who highlighted an apparent paradox in quantum mechanics. Nearly three decades later, Bell introduced a quantitative and experimentally testable framework for distinguishing quantum correlations from local classical descriptions \cite{bell1964reprinted}. Bell inequalities therefore offer an experimentally motivated route to probing whether gravity itself has a quantum nature. A widely used form is the Clauser--Horne--Shimony--Holt (CHSH) Bell inequality, where the corresponding CHSH operator is given by \cite{clauser1969proposed,horodecki1995violating}
	\begin{equation}
		\hat{B}
		=
		\vec{a}\cdot\vec{\sigma}\otimes(\vec{b}+\vec{b}')\cdot\vec{\sigma}
		+
		\vec{a}'\cdot\vec{\sigma}\otimes(\vec{b}-\vec{b}')\cdot\vec{\sigma},
	\end{equation}
where \(\vec{a},\vec{a}',\vec{b},\vec{b}'\) are unit vectors and \(\vec{\sigma} = (\sigma_x, \sigma_y, \sigma_z)\) is the vector of Pauli matrices. The associated Bell--CHSH inequality is
	\begin{equation}
		|\langle \hat{B}\rangle|
		=
		|\mathrm{tr}(\rho\hat{B})|
		\le 2.
	\end{equation}
	A violation, $|\langle \hat{B}\rangle|>2$, signals the presence of quantum correlations in the state and thus provides an experimentally appealing test of quantum features of gravity. \\
Following the formulation by Wootters \cite{Wootters1998}, the concurrence quantifies the entanglement of a two-qubit density matrix \(\rho\). This measure is derived from the spin-flipped state \(\bar{\rho} = (\sigma_y \otimes \sigma_y) \rho^\star (\sigma_y \otimes \sigma_y)\). The eigenvalues \(\lambda_i\) of the matrix \(R = \sqrt{\sqrt{\rho} \bar{\rho} \sqrt{\rho}}\) are calculated and ordered such that \(\lambda_0 \geq \lambda_1 \geq \lambda_2 \geq \lambda_3\). The concurrence is then given by
\begin{equation}
	\mathcal{C} = \text{max}\{0, \lambda_0 - \lambda_1 - \lambda_2 - \lambda_3\}.
\end{equation}
For the two-qubit state considered in this work, given by Eq.~\eqref{eq:evolved-state}, this general expression reduces to 
\begin{equation}
	\mathcal{C}
	=
	4r_1r_2t_1t_2
	\left|\sin\left(\frac{\Delta\phi}{2}\right)\right|,
	\label{eq:concurrence}
\end{equation}
	where \(\Delta\phi = \Phi_{uu} + \Phi_{dd} - \Phi_{ud} - \Phi_{du}\) is the entanglement phase. For balanced beam splitters, $r_i=t_i=1/\sqrt{2}$, this reduces to $\mathcal{C}=|\sin(\Delta\phi/2)|$. Maximal entanglement, $\mathcal{C}=1$, occurs when $\Delta\phi$ is an odd multiple of $\pi$, whereas entanglement vanishes for even multiples. For the two-qubit pure state considered in this work, the experimentally accessible Bell parameter is related to the concurrence through the following expression \cite{ghosh2001mixedness,verstraete2001entanglement}
	\begin{equation}
		B_{\max}
		=
		2\sqrt{1+\mathcal{C}^2},
		\label{eq:bell-param}
	\end{equation}
	where $B_{\max}>2$ signifies a violation of the CHSH inequality. Accordingly, we consider $B_{\max}-2$ to quantify the degree of nonlocality in our analysis. \\
	Both entanglement measures, Eqs.~\eqref{eq:concurrence} and~\eqref{eq:bell-param}, depend critically on the entanglement phase. For currently accessible mass scales and the smallest possible separations where gravity still dominates over background forces, this phase is extremely difficult to accumulate, and overcoming these experimental challenges is a major focus of current research \cite{margalit2021realization,bose2025massive}. Because different geometries yield significantly different phase accumulations, determining \(\Delta\phi\) for each configuration and identifying the geometry that maximizes it is the central task of this work. Here, we demonstrate that full geometrical optimization is indispensable for achieving this goal. In the following sections, we derive exact analytical expressions for \(\Delta\phi\) in the Stern--Gerlach and Mach--Zehnder geometries, investigate the optimal operation of each in detail, and compare all five configurations on an equal footing.

	\section{STERN--GERLACH INTERFEROMETER GEOMETRIES}
	\label{sec:sg}
	
	The Stern--Gerlach apparatus is historically one of the most foundational setups in quantum mechanics \cite{sakurai1994modern}. Since the landmark 1922 experiment by Stern and Gerlach, which provided the first direct evidence of spatial quantization of the magnetic moment, the SG configuration has become a cornerstone for matter-wave interferometry. In these devices, the spin degree of freedom is utilized via magnetic field gradients to coherently split and recombine the wavepacket of a massive particle. In the context of probing the quantum nature of gravity, such spin-based matter-wave interferometers have recently gained renewed interest. In particular, proposals by S.~Bose and collaborators have highlighted the potential of using the spins of nitrogen-vacancy (NV) centers in nanodiamonds to create macroscopic spatial superpositions \cite{bose2017spin,bose2025massive}. These systems offer a promising and experimentally viable platform for testing gravitationally induced entanglement between massive quantum systems.
	
	Building on this motivation, we now define the specific geometric configurations analyzed in this work. In an SG interferometer, the two paths are separated by an angle \(2\theta\), with a vertical displacement \(\Delta x\) between them. The horizontal arm has length \(a\), and the total path length is \(l = a + \Delta x/\sin\theta\). The two identical interferometers are separated by a distance \(d\).
	
	\subsection{Parallel-Arms Configuration}
	\label{sec:sg-parallel}
	
\begin{figure}[t]
		\centering
		\includegraphics[width=3.1in]{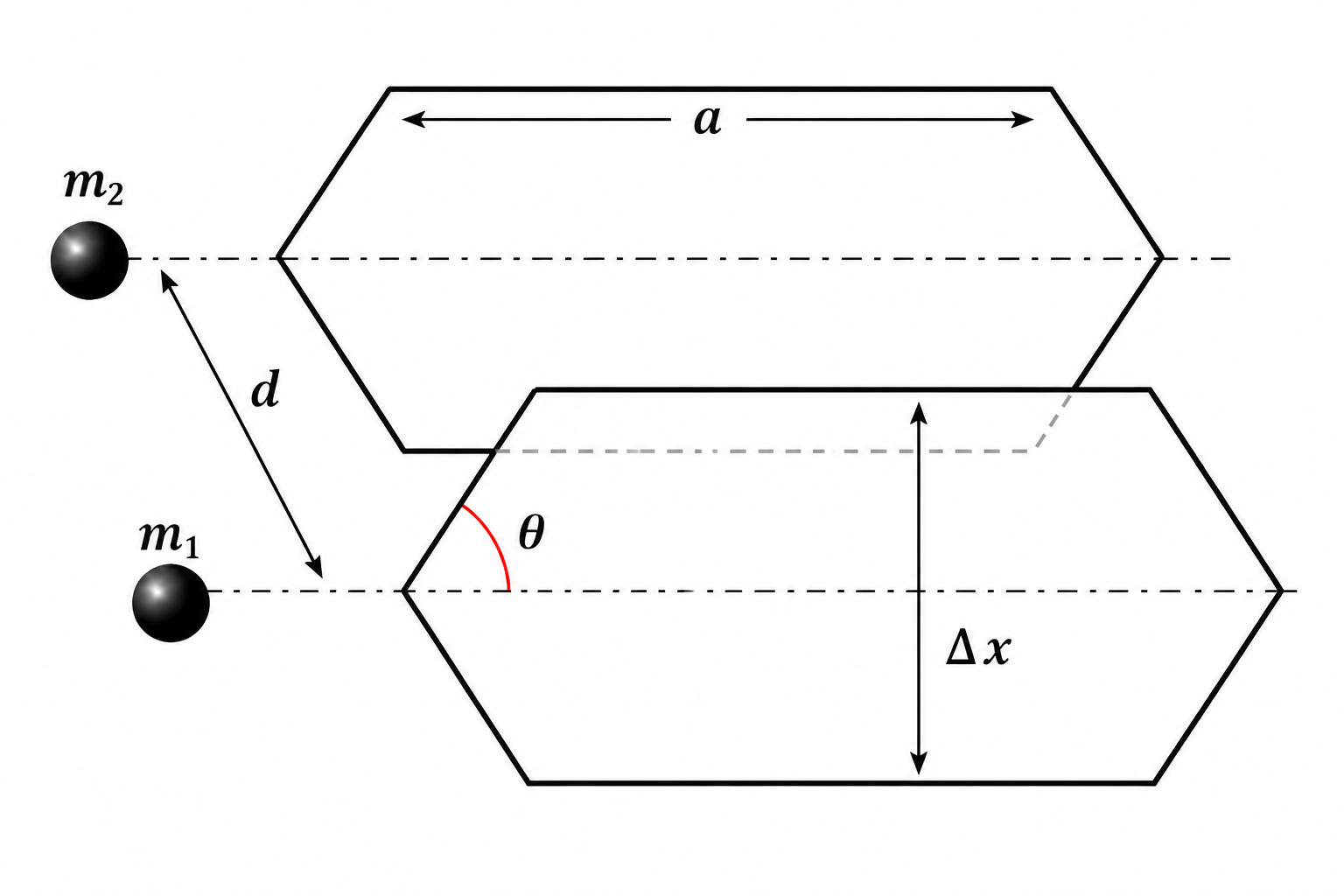}
		\caption{Schematic representation of the parallel-arms Stern--Gerlach configuration. The two interferometers are arranged such that each lies in a parallel plane separated by a distance \(d\) from the other, with all corresponding arms remaining parallel along their entire length.}
		\label{fig:sg-parallel}
	\end{figure}

	In the parallel-arms configuration, the two interferometers lie in parallel planes, with all four corresponding arms arranged parallel to each other and the two setups separated by a uniform transverse distance \(d\), as depicted in Fig.~\ref{fig:sg-parallel}. Using Eq.~\eqref{COWph}, for different paths, the corresponding gravitational phases are given by
	\begin{align}
		\Phi_{uu}/\kappa = \Phi_{dd}/\kappa
		&= \frac{a+\Delta x/\sin\theta}{d},
		\label{eq:sg-parallel-uu} \\
		\Phi_{ud}/\kappa = \Phi_{du}/\kappa
		&=
		\frac{1}{\sin\theta}
		\sinh^{-1}\left(\frac{\Delta x}{d}\right)
		+\frac{a}{\sqrt{d^2+\Delta x^2}},
		\label{eq:sg-parallel-ud}
	\end{align}
	with the phase difference \(\Delta\phi = 2\left(\Phi_{uu}-\Phi_{ud}\right)\). We assume that both particles traverse their respective interferometers with a constant speed \(v\). The common parameter
	\begin{equation}
		\kappa = \frac{G m_1 m_2}{\hbar v}
		\label{eq:kappa}
	\end{equation}
	then sets the overall scale of the gravitational phase. We express phase differences in units of \(\kappa\) to isolate the geometric factor \(\Delta\phi/\kappa\), allowing for a direct, scale-independent comparison across geometries. \\
Several recent studies have investigated this parallel configuration for probing the quantum nature of gravity, analyzing its performance under decoherence \cite{nguyen2020entanglement,tilly2021qudits,schut2023relaxation,braccini2023large}. However, a systematic optimization of its geometric parameters to maximize the accumulated phase has not yet been performed. It is therefore essential to determine the precise geometric conditions under which the combination of these individual phase contributions yields the maximum gravitationally induced phase accumulation.
	
	\subsection{Linear Configuration}
	\label{sec:sg-linear}
	
\begin{figure}[t]
	\centering
	\includegraphics[width=3.1in]{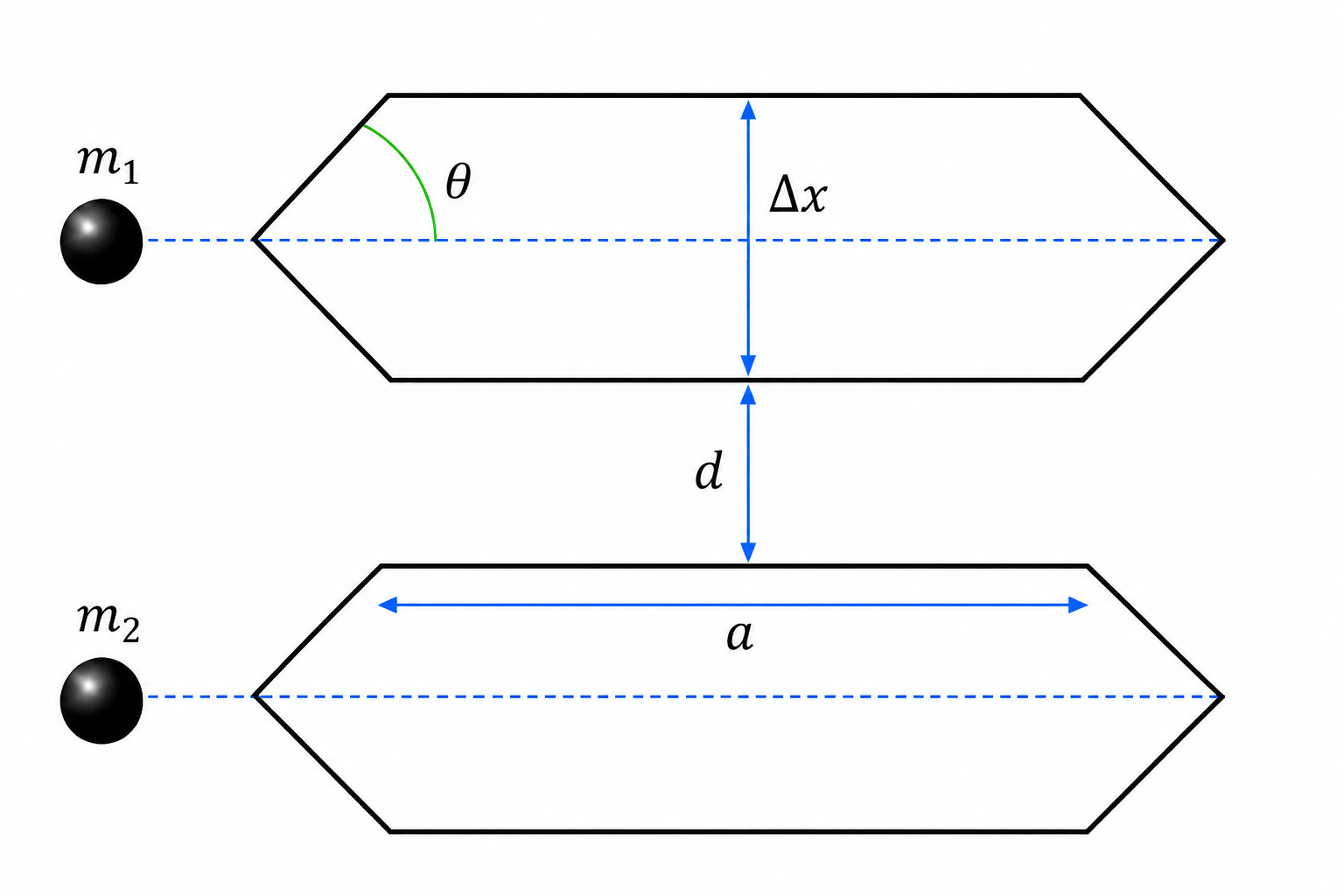}
	\caption{Schematic of the linear Stern--Gerlach configuration. The two interferometers lie in a single plane, one above the other, with their inner arms separated by a distance \(d\).}
	\label{fig:sg-linear}
\end{figure}

In the linear configuration, the two interferometers lie in the same plane, with their inner arms separated by a distance \(d\), as illustrated in Fig.~\ref{fig:sg-linear}. This linear geometry represents the original Stern--Gerlach setup proposed for the detection of gravitationally induced entanglement \cite{bose2017spin,marletto2017gravitationally} and has been extensively studied in the literature. In this work, we provide a detailed comparative analysis of the linear and parallel-arms configurations. We explicitly evaluate their maximum achievable phase accumulation alongside their robustness to geometric imperfections, such as fabrication tolerances in the path lengths. Using Eq.~\eqref{COWph}, and including full particle trajectories, the exact phases are given by 
	\begin{align}
		\Phi_{uu}/\kappa &= \Phi_{dd}/\kappa = 
		\left(\frac{\Delta x}{\sin\theta}+a\right)
		\frac{1}{d+\Delta x},
		\label{eq:sg-phi1} \\
		\Phi_{ud}/\kappa &= 
		\frac{1}{\sin\theta}
		\ln\left(1+\frac{\Delta x}{d+\Delta x}\right)
		+\frac{a}{d+2\Delta x}, \\
		\Phi_{du}/\kappa &= 
		\frac{1}{\sin\theta}
		\ln\left(1+\frac{\Delta x}{d}\right)
		+\frac{a}{d}.
		\label{eq:sg-phi3}
	\end{align}
	The entanglement phase is \(\Delta\phi = 2\Phi_{uu}-\Phi_{ud}-\Phi_{du}\). \\
Having established the exact analytical expressions for both the linear and parallel geometries, we can now systematically evaluate and compare their respective efficiencies in accumulating a gravitationally induced phase. This detailed comparative analysis is presented in the following subsection.
	
	\subsection{Phase Comparison across Stern--Gerlach Configurations}
	\label{sec:sg-optimal}
	
\begin{figure}[t]
		\centering
		\includegraphics[width=0.49\textwidth]{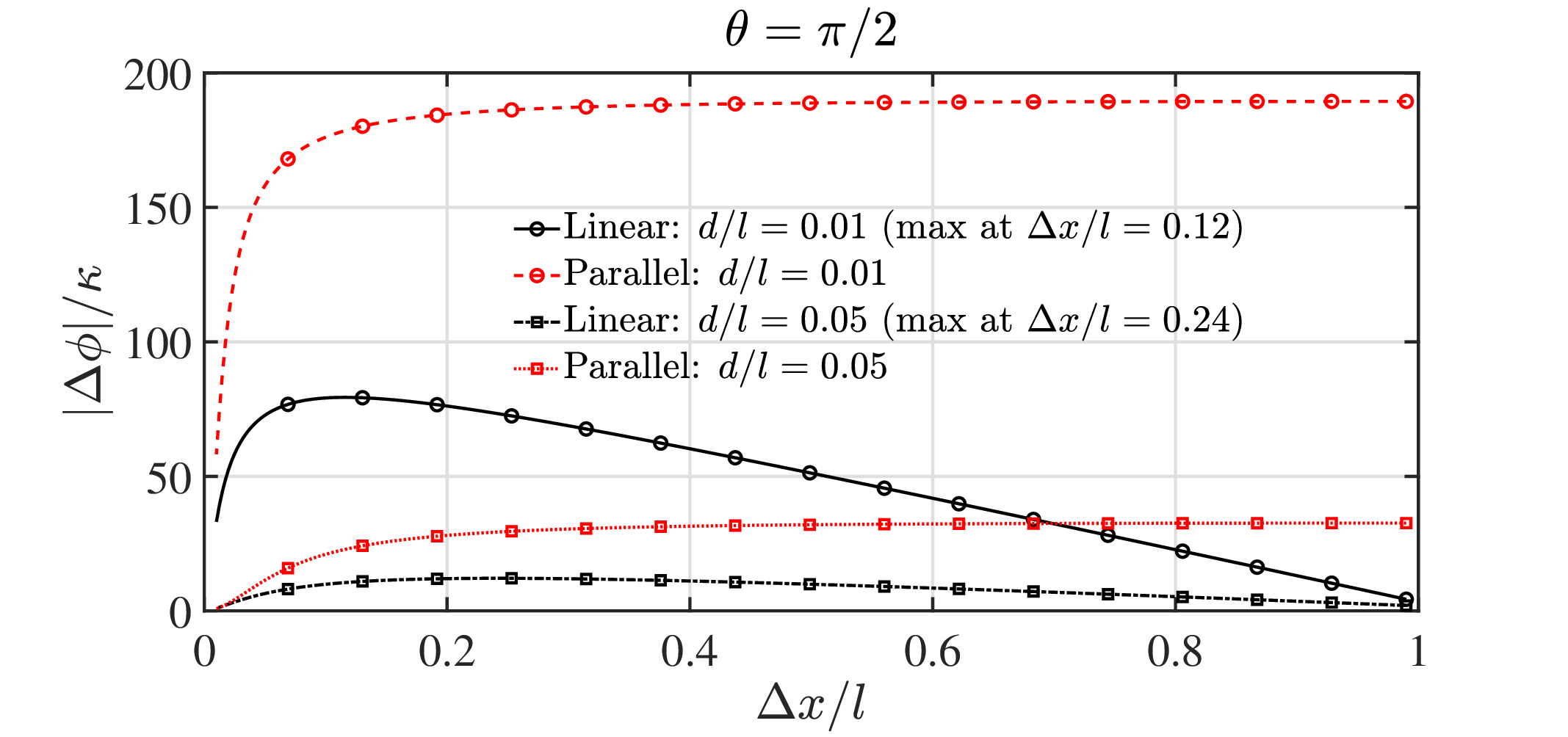}
		\caption{Rescaled phase \(|\Delta\phi|/\kappa\) versus normalized vertical displacement \(\Delta x/l\) for the SG configurations at \(\theta=\pi/2\). The linear configuration is represented by the solid curve for \(d/l=0.01\) and the dash-dotted curve for \(d/l=0.05\). The parallel configuration is represented by the dashed curve for \(d/l=0.01\) and the dotted curve for \(d/l=0.05\). The parallel geometry consistently produces a larger maximum phase, peaking as \(\Delta x/l \to 1\), while the linear geometry exhibits a nonmonotonic peak at lower \(\Delta x/l\).}
		\label{fig:sg-phase-dx}
	\end{figure}

	Figure~\ref{fig:sg-phase-dx} illustrates the rescaled phase \(|\Delta\phi|/\kappa\) as a function of the normalized vertical displacement \(\Delta x/l\) for a fixed deviation angle of \(\theta=\pi/2\). The solid curves correspond to the linear configuration at \(d/l=0.01\), while the dash-dotted curves correspond to the linear configuration at \(d/l=0.05\). The dashed curves correspond to the parallel configuration at \(d/l=0.01\), and the dotted curves correspond to the parallel configuration at \(d/l=0.05\). As \(\Delta x/l \to 0\), the accumulated phase vanishes entirely because the spatial superposition between the two paths collapses. Furthermore, the phase variation is abrupt in this regime. For instance, for vertical displacements smaller than the linear phase maximum, the phase exhibits a sharp decrease. This strong geometric dependence demonstrates that the widely used negligible \(\Delta x\) approximation does not yield an accurate description of the accumulated phase.  As illustrated in Fig.~\ref{fig:sg-phase-dx}, the parallel configuration also yields a larger phase across nearly the entire parameter space. The two configurations become comparable only in the regime \(\Delta x/l \lesssim 0.02\) for \(d/l = 0.05\), where both interferometers become inefficient at accumulating a significant phase. \\
For the parallel setup, the phase increases monotonically as \(\Delta x/l \to 1\). In this limit, Eqs.~\eqref{eq:sg-parallel-uu} and~\eqref{eq:sg-parallel-ud} yield a constant asymptotic value of \(\Delta\phi/\kappa \simeq 2l/d - 2\sinh^{-1}(l/d)\). Since this dominant term is independent of \(\Delta x\), the phase profile plateaus for \(\Delta x/l > 0.2\), as shown in Fig.~\ref{fig:sg-phase-dx}. Consequently, the experiment is largely insensitive to precise tuning of the vertical displacement.
In contrast, the linear configuration is nonmonotonic. It reaches its maximum phase at specific, intermediate values of \(\Delta x/l\) before decreasing. From Fig.~\ref{fig:sg-phase-dx}, these optimal points are approximately \((\Delta x/l)_{\text{opt}} \approx 0.12\) for \(d/l=0.01\) and \((\Delta x/l)_{\text{opt}} \approx 0.24\) for \(d/l=0.05\).
	
The enhancement provided by the parallel geometry over the linear one is substantial. When the parallel configuration is evaluated at its asymptotic limit (\(\Delta x/l \to 1\)), it yields an enhancement factor of approximately \(2.4\) relative to the peak phase of the linear configuration at \(d/l=0.01\), and \(2.7\) relative to the peak phase of the linear configuration at \(d/l=0.05\).
	
\begin{figure}[t]
		\centering
		\includegraphics[width=0.49\textwidth]{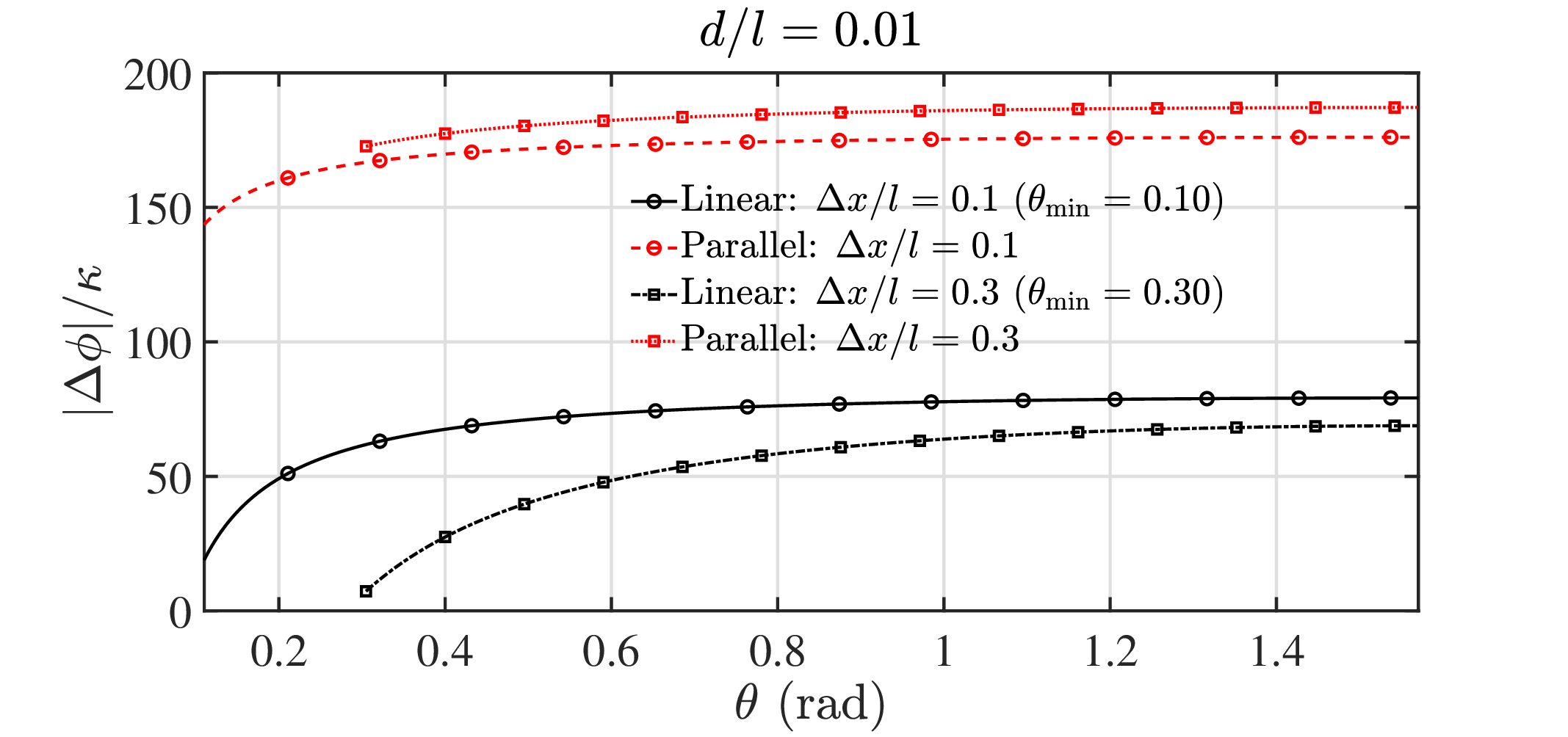} 
		\caption{Rescaled phase \(|\Delta\phi|/\kappa\) as a function of the deviation angle \(\theta\) for different normalized vertical displacements \(\Delta x/l\). The phase is maximized at \(\theta=\pi/2\), reflecting the optimal geometric projection of the interferometer arms along the gravitational gradient.}
		\label{fig:sg-phase-theta}
	\end{figure}

 Figure~\ref{fig:sg-phase-theta} displays the rescaled phase as a function of \(\theta\) for representative values of \(\Delta x/l\). The numerical evaluation demonstrates that, regardless of the chosen displacement, the phase consistently reaches a global maximum exactly at \(\theta=\pi/2\). At this angle, the vertical projection \(\Delta x\) is maximized for the given path length \(l\), thereby optimizing the signal. Consequently, \(\theta=\pi/2\) represents the optimal angular configuration for the experiment.
	
	\subsection{Optimal Stern--Gerlach Geometry}
	\label{sec:sg-optimal-geometry}
	
\begin{figure}[t]
		\centering
		\includegraphics[width=\columnwidth]{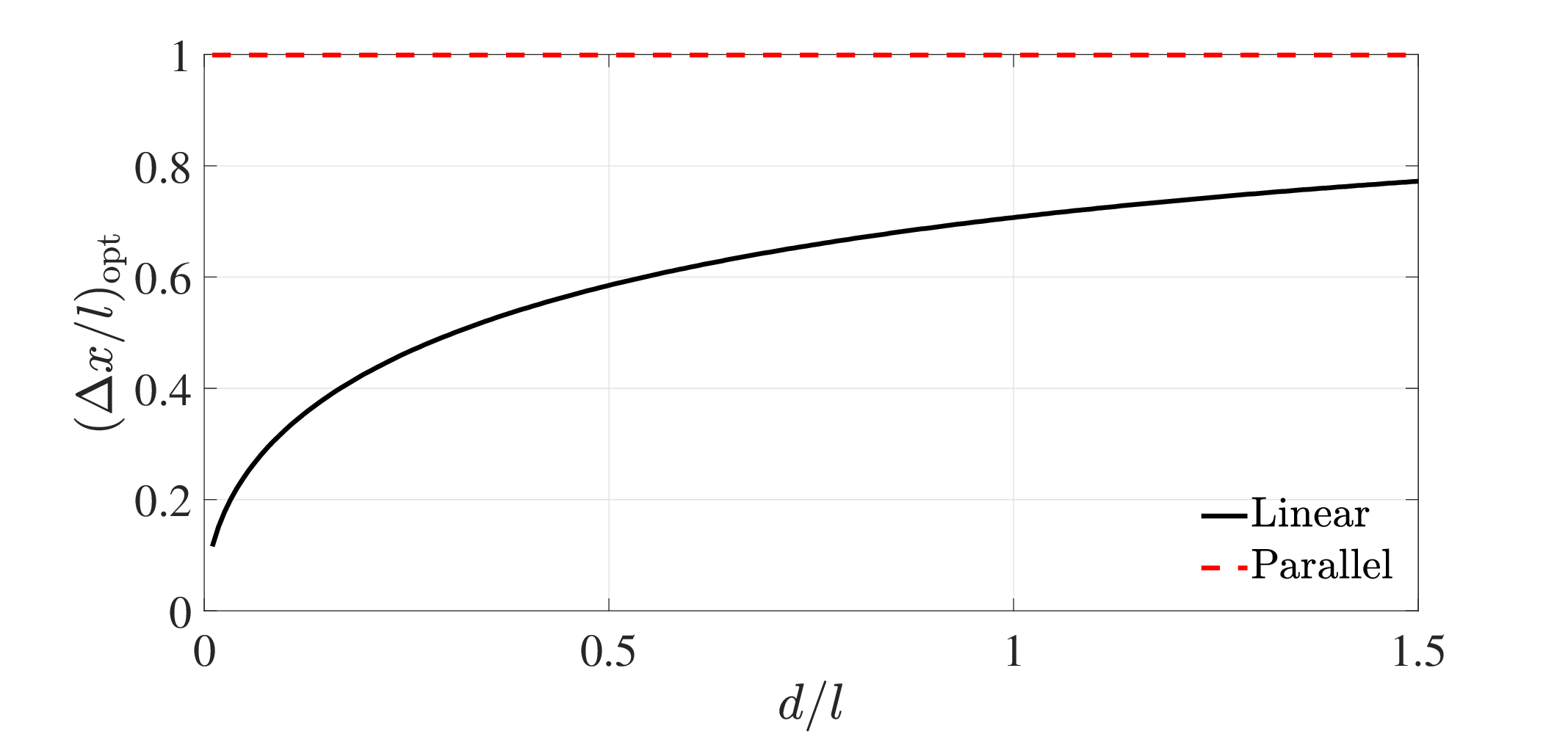}
		\caption{Optimal normalized vertical displacement \((\Delta x/l)_{\text{opt}}\) that maximizes the absolute rescaled phase \(|\Delta\phi|/\kappa\) as a function of the normalized interferometer separation \(d/l\) for the SG configurations. The solid curve represents the linear configuration, while the dashed line represents the parallel configuration, which remains constant across the entire parameter space.}
		\label{fig:sg-opt}
	\end{figure}

	The nonmonotonic behavior of the linear setup indicates the existence of an optimal geometric configuration for each choice of the separation \(d/l\).  To illustrate this behavior, Fig.~\ref{fig:sg-opt} plots the optimal geometric ratio \((\Delta x/l)_{\text{opt}}\) as a function of \(d/l\). The solid  curve corresponds to the linear configuration, which exhibits a monotonic increase. The dashed  line corresponds to the parallel configuration, which remains fixed at its maximum geometric limit across the entire parameter space. In practice, the minimum achievable separation is bounded by non-gravitational forces such as van der Waals interactions \cite{maleki2026role}, and the optimal value of \(\Delta x/l\) must be selected according to the experimentally accessible range of \(d/l\).
	
	While the linear configuration requires this careful balance between \(d/l\) and \(\Delta x/l\), the parallel geometry offers a compelling simplification. As shown in Fig.~\ref{fig:sg-phase-dx}, the parallel phase does not exhibit a local maximum at intermediate values. Instead, it rises monotonically and plateaus as \(\Delta x/l \to 1\). This means that the optimal configuration for the parallel setup is universally approached at the geometric limit where the vertical arm is maximized. This structural difference is a key advantage of the parallel design. It eliminates the need to fine-tune the vertical displacement to a specific, separation-dependent value. The experiment can simply be operated at the largest practically achievable \(\Delta x/l\), ensuring that the system remains in the high-phase regime regardless of the exact value of \(d/l\). 
	\section{MACH--ZEHNDER INTERFEROMETER GEOMETRIES}
	\label{sec:mz}
Beyond the Stern--Gerlach configurations analyzed in the previous section, Mach--Zehnder (MZ) interferometers offer another prominent platform for generating and probing macroscopic spatial superpositions. Owing to their distinct operational mechanism, MZ geometries have gained significant attention in the context of gravitationally induced entanglement proposals \cite{bose2017spin,marletto2017gravitationally}. However, just as with the SG setups, the optimal geometric configuration of an MZ interferometer is a crucial yet often under-explored factor in maximizing the observable quantum signature. We now apply the formalism to MZ interferometers, which are characterized by the horizontal arm length \(a\), vertical arm separation \(b\), and separation \(d\). 
\subsection{Parallel-Arms Configuration}
\label{sec:mz-parallel}
\begin{figure}[t]
	\centering
	\includegraphics[width=3.5in]{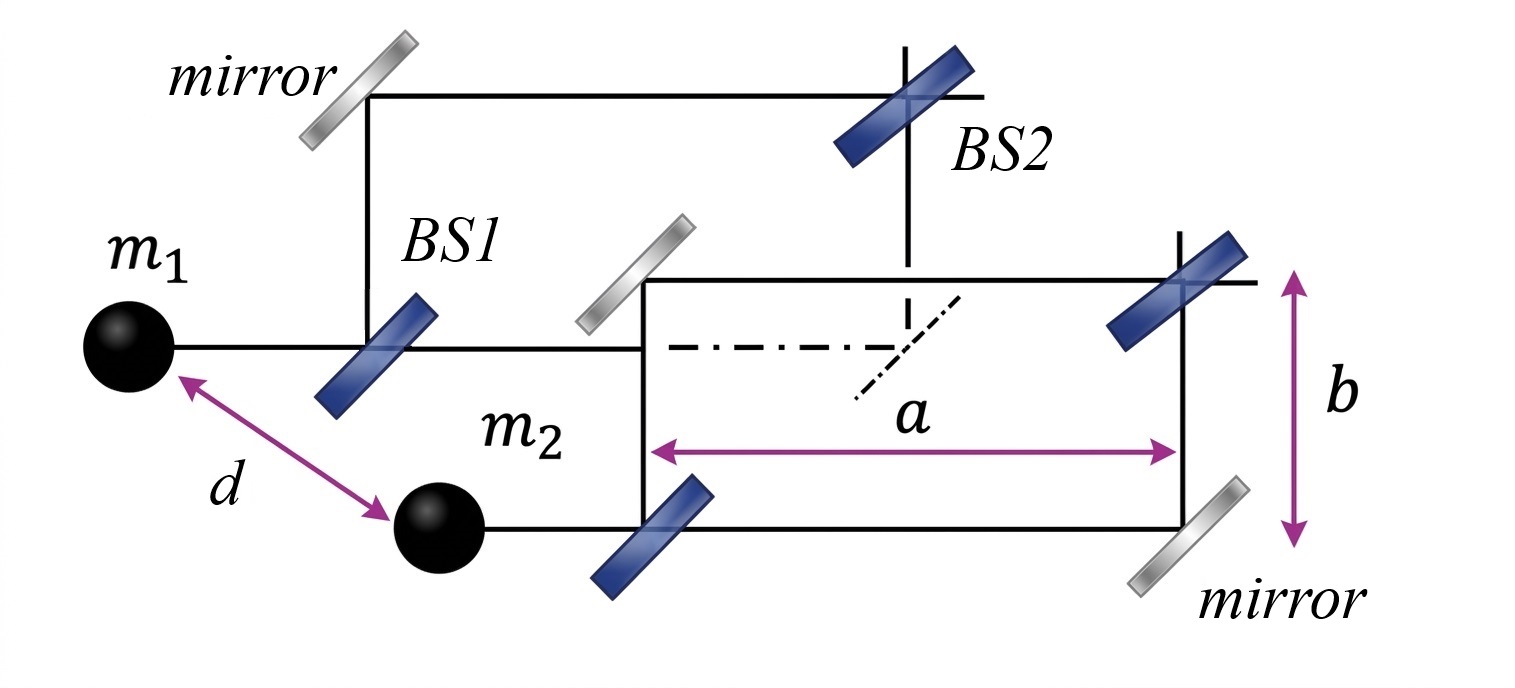}
	\caption{Schematic representation of the Mach--Zehnder interferometer in the parallel-arms (PAMZ) configuration. The setup consists of two identical MZ interferometers lying in parallel planes separated by a distance \(d\), with all corresponding vertical and horizontal arms remaining parallel along their entire length. The horizontal arm length is \(a\) and the vertical separation is \(b\).}
	\label{fig:mz-parallel}
\end{figure}
While linear MZ configurations have been widely studied in the context of testing gravitationally induced entanglement, the parallel-arms MZ (PAMZ) geometry has received comparatively little attention. Here, we demonstrate that this specific configuration possesses unique phase accumulation properties. Similar to the parallel SG setup, the two MZ interferometers lie in parallel planes separated by a distance \(d\), with all corresponding arms remaining parallel along their entire length, as shown in Fig.~\ref{fig:mz-parallel}. The explicit phase expressions are given by
	\begin{equation}
		\begin{aligned}
			\Phi_{uu}/\kappa &= \Phi_{dd}/\kappa=\frac{l}{d}, \\
			\Phi_{ud}/\kappa &= \Phi_{du}/\kappa = 
			\begin{cases} 
				h(b,d) + \frac{l-2b}{\sqrt{d^2+2b^2}}, & l > 2b, \\[2ex]
				h(l-b,d) + \frac{2b-l}{\sqrt{d^2+2(l-b)^2}}, & \text{otherwise},
			\end{cases}
		\end{aligned}
		\label{eq:mz-pa-phases}
	\end{equation}
	where \(h(x,y) = \sqrt{2} \ln \left( (\sqrt{2}x + \sqrt{y^2+2x^2})/y \right)\). The entanglement phase is \(\Delta\phi = 2(\Phi_{uu}-\Phi_{ud})\).
	Beyond this parallel configuration, the MZ setup admits other possible geometries. In the following subsections, we evaluate how two distinct linear orientations modify the phase accumulation through their specific spatial arrangements.

	\subsection{Linear Configuration with Translational Symmetry}
\label{sec:mz-translational}
\begin{figure}[t]
		\centering
		\includegraphics[width=3in]{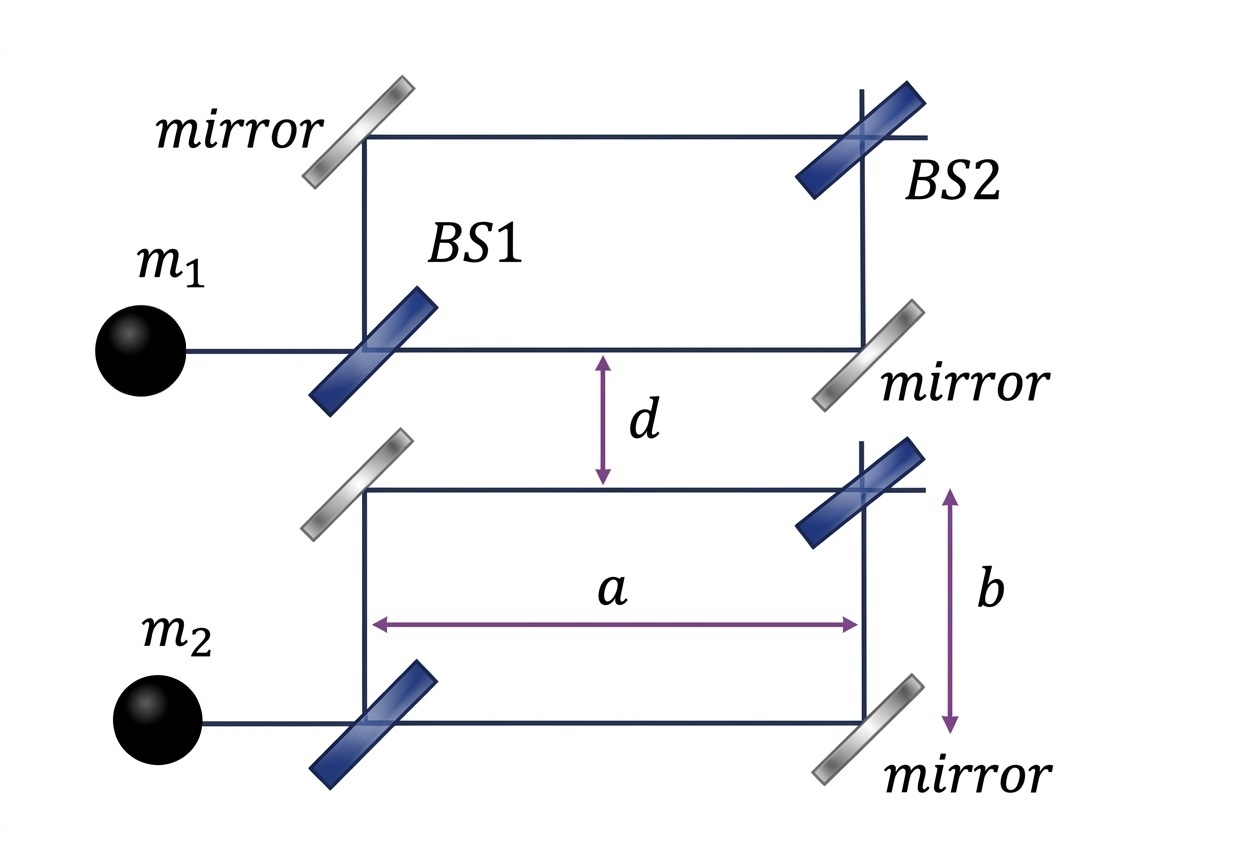}
		\caption{Schematic of the Mach--Zehnder interferometer with translational symmetry (TSMZ). The two interferometers are identical in shape and can be mapped onto one another by a displacement of \(d+b\) along the vertical direction.}
		\label{fig:mz-translational}
	\end{figure}

Among the linear configurations of the MZ class, we first analyze the translation-symmetric setup (TSMZ), shown in Fig.~\ref{fig:mz-translational}. In this geometry, the two interferometers can be mapped onto one another by a displacement of \(d+b\) along the vertical direction. This specific orientation corresponds to the original Mach--Zehnder setup proposed by Marletto and Vedral \cite{marletto2017gravitationally} for the detection of gravitationally induced entanglement, and it has been extensively discussed in subsequent literature. Here, we examined this configuration, paying particular attention to the often-overlooked contribution of the vertical arms. We demonstrated that improperly approximating the phase accumulation by treating the vertical paths as negligible can lead to incorrect conclusions regarding the experiment's feasibility. Specifically, the more accurately the width is chosen to fulfill this approximation condition, the smaller and more ineffective the accumulated phase becomes. By explicitly including the vertical path contributions, we avoid these pitfalls and obtain the exact phases
	\begin{align}
		\Phi_{uu}/\kappa &= \Phi_{dd}/\kappa = \frac{l}{d+b}, \label{eq:mz-ts-uu} \\
		\Phi_{ud}/\kappa &=
		\begin{cases}
			2g(b) + \frac{l-2b}{\sqrt{d^2+b^2}}, & b \leq l/2, \\[2ex]
			2g(l-b) +\frac{2b-l}{\sqrt{(d+2b-l)^2+(l-b)^2}}, & \text{otherwise},
		\end{cases} \label{eq:mz-ts-ud} \\
		\Phi_{du}/\kappa &=
		\begin{cases}
			2f(b,d+b) + \frac{l-2b}{\sqrt{(d+2b)^2+b^2}}, & b \leq l/2, \\[2ex]
			2f(l-b,d+b) +\frac{2b-l}{\sqrt{(d+l)^2+(l-b)^2}}, & \text{otherwise},
		\end{cases} \label{eq:mz-ts-du}
	\end{align}
	where the auxiliary functions are
	\begin{align}
		f(x, y) &= \frac{1}{\sqrt{2}} \ln \left( \frac{2x + y + \sqrt{2(x + y)^2 + 2x^2}}{y(1 + \sqrt{2})} \right), \label{eq:mz-ts-f} \\ 
		g(x) &= \frac{1}{\sqrt{2}} \Bigg( \sinh^{-1}\left(\frac{2x}{d + b} - 1\right) + \sinh^{-1}\left( 1\right) \Bigg). \label{eq:mz-ts-g}
	\end{align}
	The entanglement phase is \(\Delta\phi = 2\Phi_{uu}-\Phi_{ud}-\Phi_{du}\). 
	The exact analytical phases for the TSMZ configuration are fully specified by the expressions above. The auxiliary functions \(f\) and \(g\) capture the non-trivial contributions from the vertical segments, which depend on the specific horizontal arrangement of the two interferometers. These exact expressions reveal features that are not apparent in previous studies that relied on geometric approximations. We examine these features in detail in the following analysis.
	
	\subsection{Linear Configuration with Mirror Symmetry}
	\label{sec:mz-mirror}
	
\begin{figure}[t]
		\centering
		\includegraphics[width=2.7in]{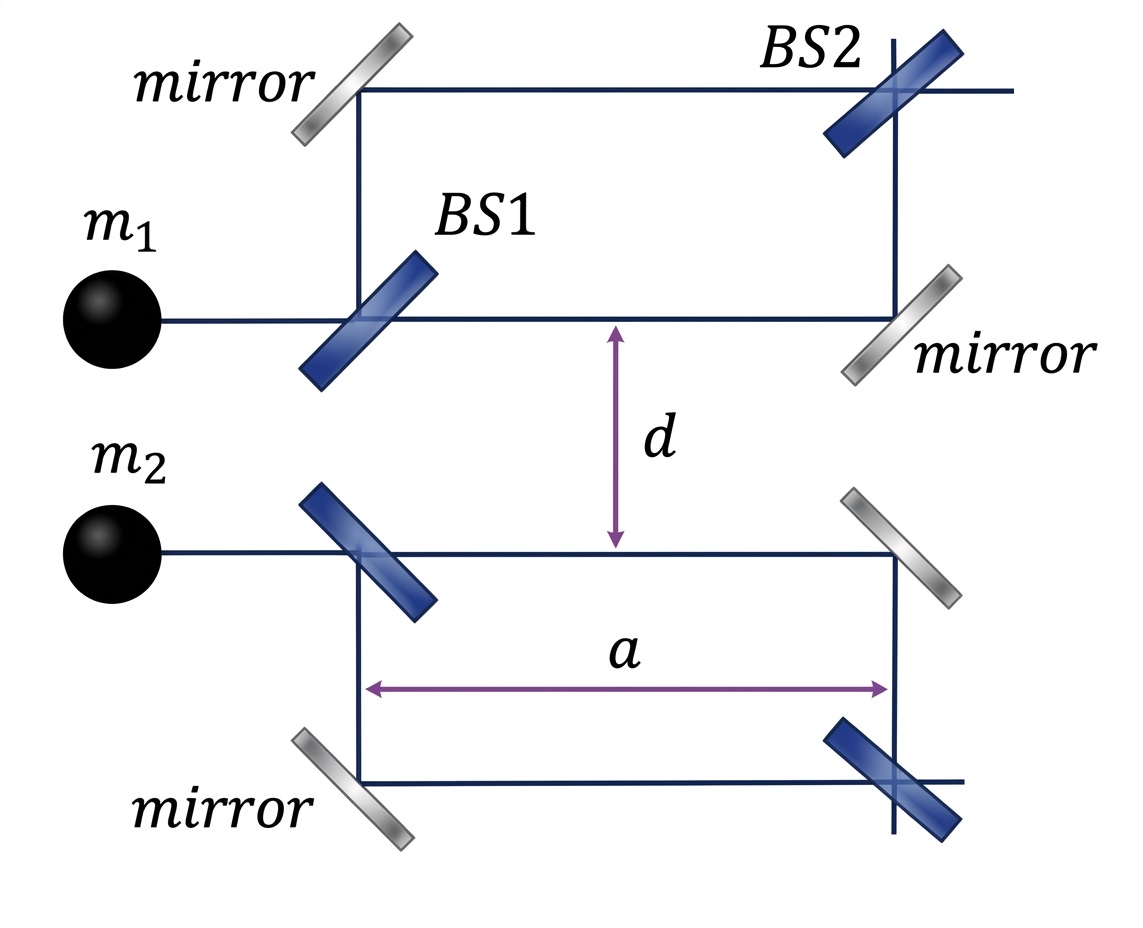}
		\caption{Schematic of the Mach--Zehnder interferometer with mirror symmetry (MSMZ). This arrangement makes the two interferometers exact mirror images across the horizontal mid-plane.}
		\label{fig:mz-mirror}
	\end{figure}

We conclude our analysis of the Mach--Zehnder geometries with the mirror-symmetric setup (MSMZ), shown in Fig.~\ref{fig:mz-mirror}. Here, we consider the exact phases and place this setup within our comparative framework to benchmark it directly against the translation-symmetric and parallel-arms configurations. This allows us to systematically evaluate its relative advantages and disadvantages with respect to phase magnitude and geometric robustness. Following the same procedure as before, we include the full particle trajectories to obtain the exact phases 
\begin{equation}
		\begin{aligned}
			&	\Phi_{uu}/\kappa = \Phi_{dd}/\kappa \\
			&= \begin{cases}
				f(b,d) + f(b,d+b) + \frac{l-2b}{\sqrt{(d+b)^2+b^2}}, & b \leq l/2, \\[2ex]
				f(l-b,d) + f(l-b,d+3b-l) + \tilde{g}(b, d, l), & \text{otherwise},
			\end{cases}
		\end{aligned}
		\label{eq:mz-ms-uu}
	\end{equation}
	where \(\tilde{g}(b,d,l)= \frac{1}{2} \ln \left( \frac{d+3b-l+\sqrt{(l-b)^2+(d+3b-l)^2}}{d+l-b+\sqrt{(l-b)^2+(d+l-b)^2}} \right)\), and the cross terms are
	\begin{align}
		\Phi_{du}/\kappa &= \frac{1}{2} \ln \left( \frac{d+2b}{d} \right) + \frac{l-b}{d}, \label{eq:mz-ms-du} \\
		\Phi_{ud}/\kappa &= \frac{1}{2} \ln \left( \frac{d+2b}{d} \right) + \frac{l-b}{d+2b}. \label{eq:mz-ms-ud}
	\end{align}
	The entanglement phase for the MSMZ geometry is then given by \(\Delta\phi = 2\Phi_{uu} - \Phi_{ud} - \Phi_{du}\). 
	
		The exact phase expressions derived above for the parallel, translation-symmetric, and mirror-symmetric geometries provide the foundation for a direct performance comparison. We now evaluate these analytical results to contrast their phase accumulation profiles and identify the most favorable configuration across the relevant parameter space.
	
	\subsection{Phase Comparison across Mach--Zehnder Configurations}
	\label{sec:mz-optimal}
	
\begin{figure}[t]
		\centering
		\includegraphics[width=0.49\textwidth]{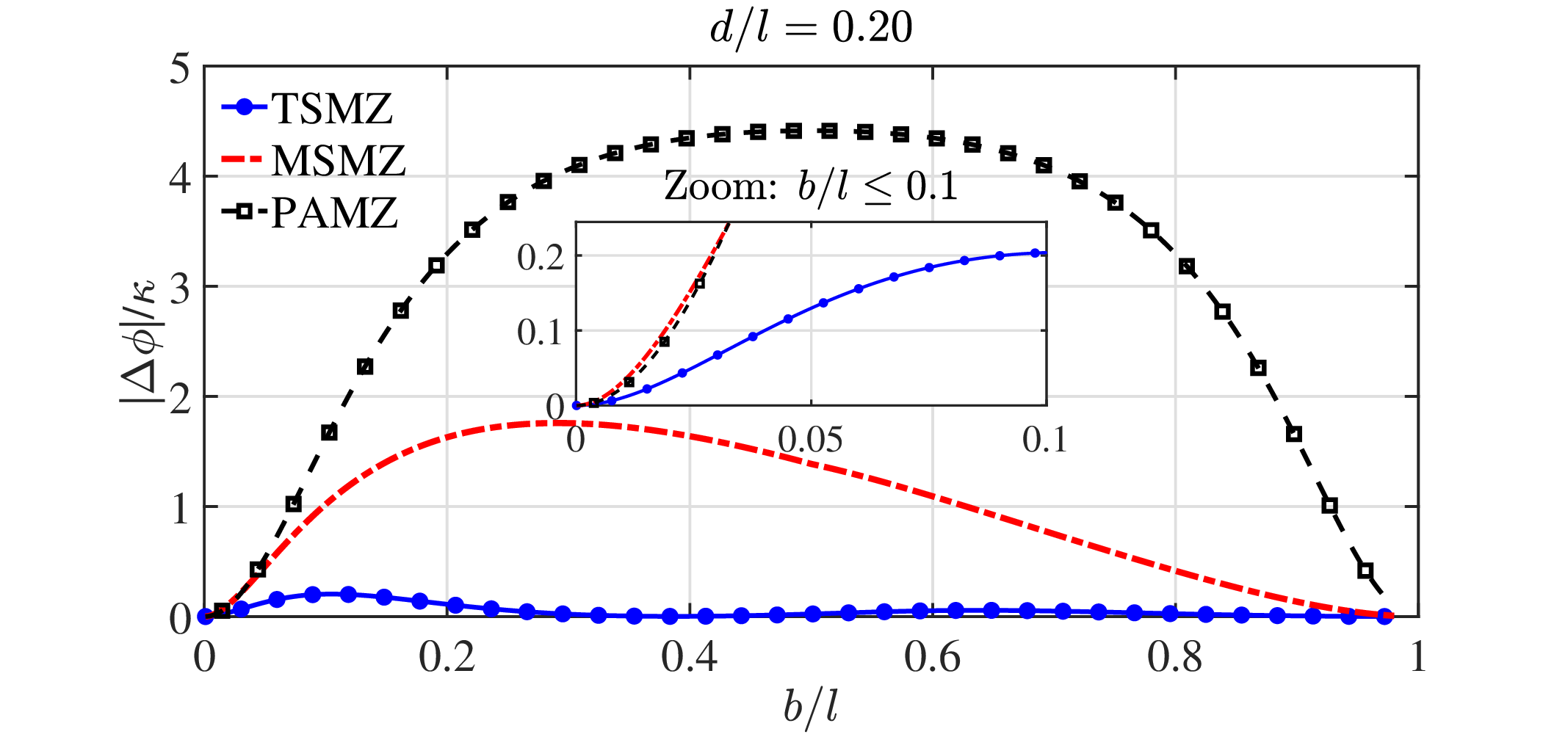}
		\caption{Rescaled phase \(|\Delta\phi|/\kappa\) versus normalized vertical arm separation \(b/l\) for the MZ configurations at \(d/l = 0.20\). The translation-symmetric (TSMZ) configuration is shown as the curve with circle markers. The mirror-symmetric (MSMZ) configuration is shown as the dash-dotted curve. The parallel-arms (PAMZ) configuration is shown as the dashed curve with square markers. The TSMZ phase drops to near zero for \(b/l \sim 0.4\), while PAMZ maintains a broad, dominant peak.}
		\label{fig:mz-phase-0p2}
	\end{figure}

\begin{figure}[t]
		\centering
		\includegraphics[width=0.49\textwidth]{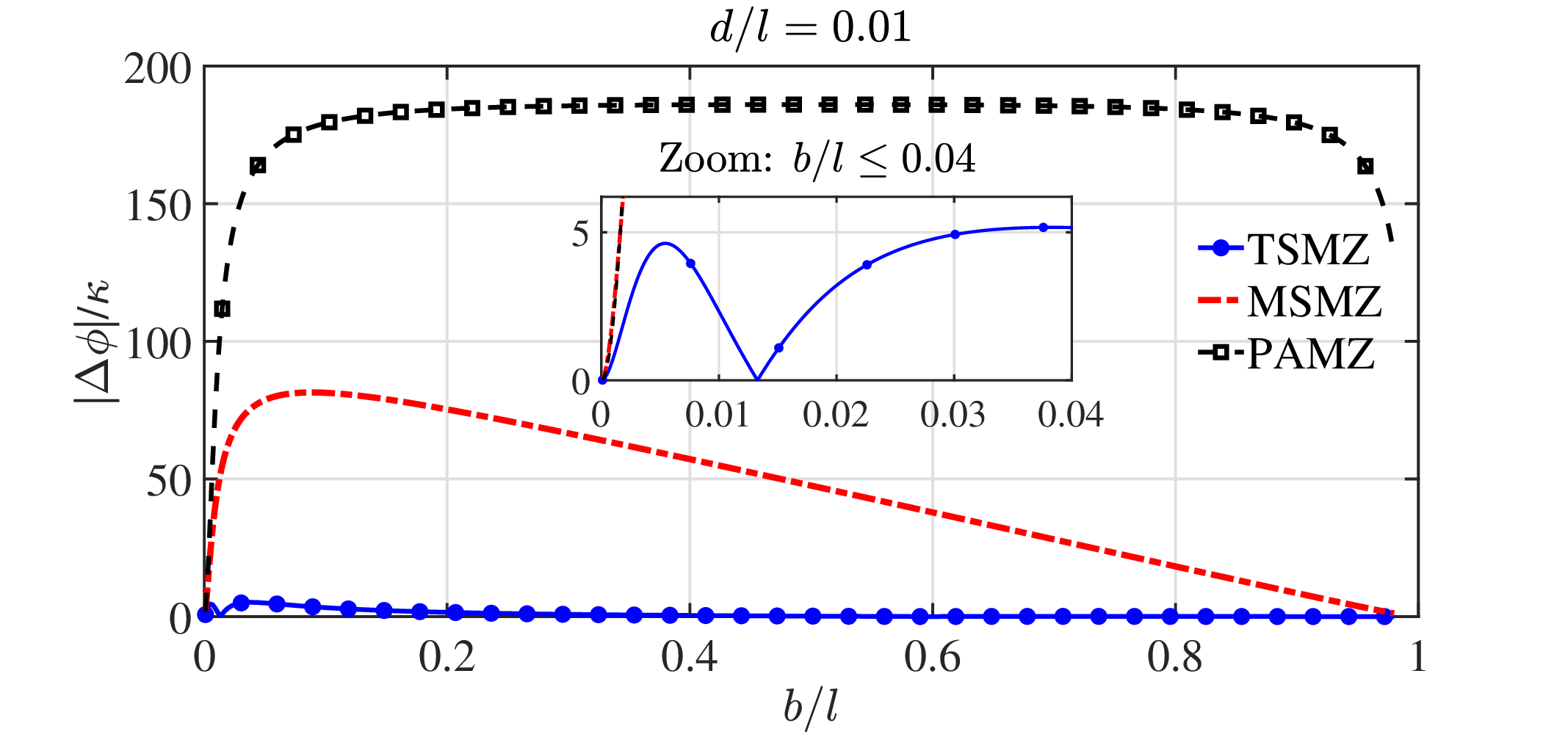}
		\caption{Rescaled phase \(|\Delta\phi|/\kappa\) versus \(b/l\) for the MZ configurations at \(d/l = 0.01\). The line and marker styles are identical to Fig.~\ref{fig:mz-phase-0p2}. The overall phase magnitude increases dramatically as the interferometers are brought closer together. The PAMZ configuration reaches a maximum phase of nearly \(190\) and maintains a broad plateau from \(b/l \approx 0.2\) to \(0.8\), demonstrating its geometric robustness.}
		\label{fig:mz-phase-0p01}
	\end{figure}

	Figures~\ref{fig:mz-phase-0p2} and~\ref{fig:mz-phase-0p01} compare the rescaled phase for \(d/l = 0.20\) and \(d/l = 0.01\). The curves with circle markers (TSMZ) exhibit exact phase nulls, dropping to zero at specific values of \(b/l\). Both the dash-dotted curves (MSMZ) and the dashed curves with square markers (PAMZ) eliminate these nulls and maintain a non-zero phase profile. Crucially, the PAMZ configuration consistently yields the largest phase. At \(d/l = 0.20\), PAMZ peaks at approximately \(4.5\), which is roughly \(2.5\) times the MSMZ peak. At \(d/l = 0.01\), PAMZ reaches approximately \(190\), far exceeding the MSMZ peak of roughly \(80\), while maintaining a broad plateau from \(b/l \approx 0.2\) to \(0.8\).
	
	The prominent flat plateau observed in the PAMZ phase profiles, and its widening as \(d/l\) decreases, can be rigorously understood by analyzing the exact phase expression in Eq.~\eqref{eq:mz-pa-phases}. This plateau is most pronounced in the central regime \(d \ll b \ll l-d\), yielding a leading-order rescaled phase of \(\Delta\phi/\kappa \sim 2l/d\). For an interaction time \(T\), this is equivalent to an absolute phase of \(\Delta\phi \sim 2Gm_1m_2T/(\hbar d)\), which is of the same order as the maximum accumulated phase of the parallel SG configuration. The physical constraints of the interferometer determine the boundaries of this plateau. Specifically, these boundaries correspond to the collapse of the spatial superposition at \(b \to 0\) and the vanishing of the horizontal arm at \(b \to l\). The phase drops sharply at these edges within boundary layers of width approximately \(d\). As \(d/l \to 0\), these boundary layers shrink to vanishingly narrow regions.
	
Conversely, the TSMZ configuration develops interior phase zeros in the small-separation regime. In the limit \(d/l\ll1\), the phase vanishes trivially at the endpoints \(b=0\) and \(b=l\), as it does for all geometries. However, it also acquires a first interior zero scaling with the separation, located at \(b\simeq1.3\,d\). For \(b>l/2\), a second interior zero appears near \(b/l\simeq0.6\) as \(d/l\to0\). As \(d/l\) increases, this second zero shifts to smaller values and approaches the first zero at \(b/l\simeq 0.4\). For \(d/l\gtrsim0.2\), these interior zeros disappear entirely. Nevertheless, the regime \(d/l \ll 1\) is generally regarded as the most experimentally relevant one, because maximizing the accumulated phase strongly favors minimizing the separation \(d\). For a typical coherence time of \(T \sim 1\,\mathrm{s}\), the path length is given by \(l = vT\). The condition \(d/l \ll 1\) is therefore typically satisfied for standard velocities.  In this context, provided that the operating regime satisfies \(d/l \lesssim 0.2\), the presence and parameter dependence of these interior zeros constitute an intrinsic experimental limitation of the TSMZ configuration. However, exceptionally slow particles could reduce \(l\) to the point where it becomes comparable to \(d\), pushing the experiment out of this favorable regime.
	
	\subsection{Optimal Mach--Zehnder Geometry}
	\label{sec:mz-optimal-geometry}
	
\begin{figure}[t]
		\centering
		\includegraphics[width=\columnwidth]{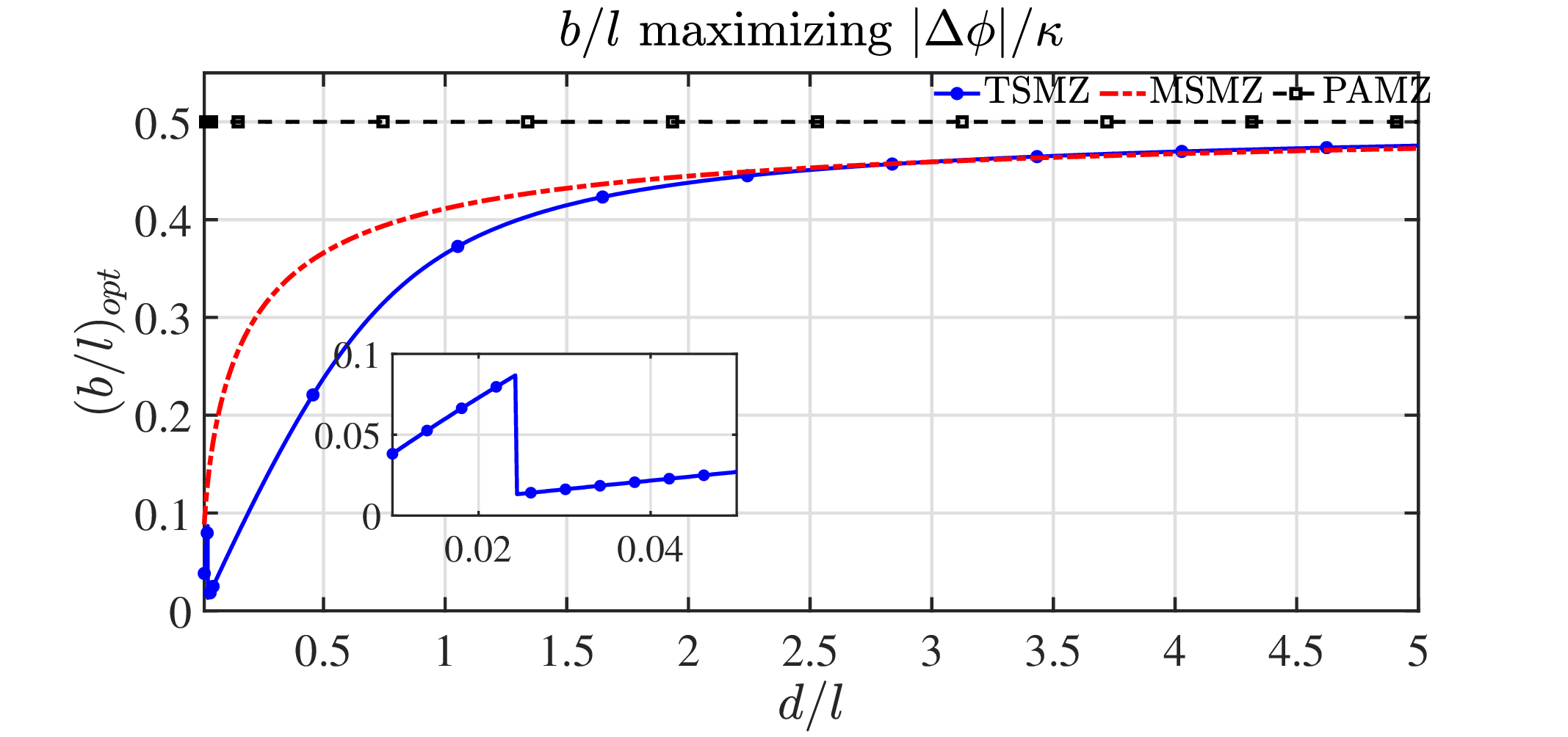}
		\caption{Optimal normalized vertical arm separation \(b/l\) that maximizes the absolute rescaled phase \(|\Delta\phi|/\kappa\) as a function of the normalized interferometer separation \(d/l\). The TSMZ setup (curve with circle markers) exhibits an abrupt discontinuity, highlighted in the inset for small separations (\(d/l \in [0.01, 0.05]\)). The MSMZ (dash-dotted curve) and PAMZ (curve with square markers) configurations remain smooth and continuous.}
		\label{fig:MZopptimum}
	\end{figure}

	To determine the most favorable operating points, Fig.~\ref{fig:MZopptimum} plots the optimal \(b/l\) that maximizes \(|\Delta\phi|/\kappa\) versus \(d/l\). The PAMZ setup (square markers) stays fixed at \(b/l = 0.5\) in all parameter regions, while the MSMZ setup (dash-dotted curve) shows a smooth, monotonic increase  toward \(0.5\). In stark contrast, the TSMZ setup (circle markers) exhibits a pathological discontinuity, highlighted in the zoomed inset of Fig.~\ref{fig:MZopptimum}. For \(d/l \approx 0.02\), the maximum lies at \(b/l \approx 0.08\). As \(d/l\) increases beyond this threshold, the optimal \(b/l\) abruptly jumps to a much smaller value. A tiny variation in \(d\) would therefore completely shift the operating point of a TSMZ interferometer, making it highly impractical in this regime.
	
	\begin{figure}[t]
		\centering
		\includegraphics[width=\columnwidth]{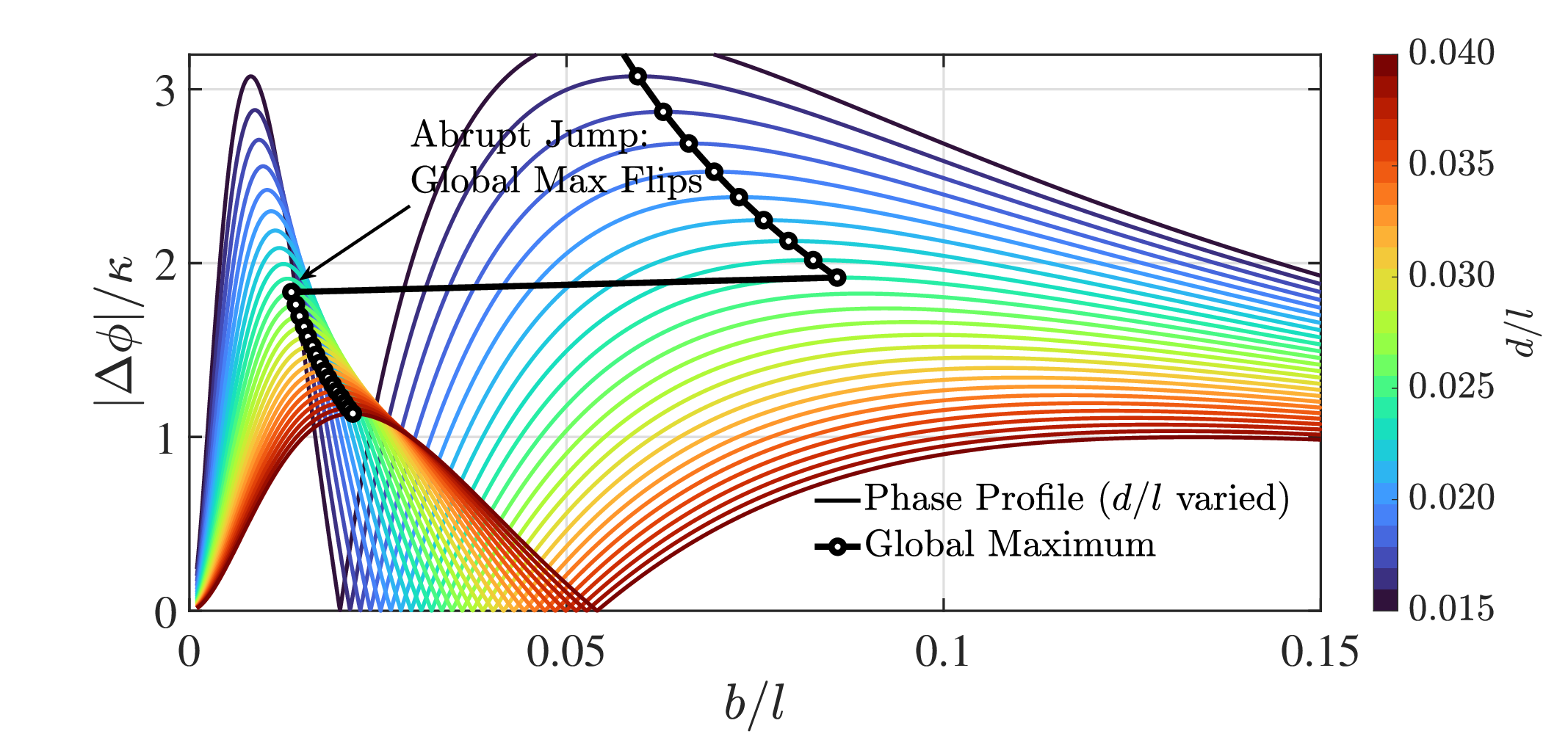}
		\caption{Evolution of the TSMZ phase profile \(|\Delta\phi|/\kappa\) as a function of the normalized vertical arm separation \(b/l\). The curves (colored in the online version) represent different values of the normalized interferometer separation \(d/l\), ranging from \(0.015\) to \(0.040\), as indicated by the colorbar. The thick curve with markers traces the location of the global maximum for each \(d/l\). For small \(d/l\) the global maximum resides on the broad right-hand peak; as \(d/l\) increases, this peak diminishes in height relative to the sharp left-hand peak, and at a critical value near \(d/l \approx 0.025\) the global maximum abruptly jumps from the right peak to the left peak, directly explaining the discontinuous optimal geometry observed in Fig.~\ref{fig:MZopptimum}.}
		\label{fig:TSMZbOphase}
	\end{figure}
	
	The discontinuity in the optimal geometry of the TSMZ setup, can be understood physically from the full phase profile across the parameter space. Figure~\ref{fig:TSMZbOphase} shows this evolution by plotting \(|\Delta\phi|/\kappa\) versus \(b/l\) for separations from \(d/l = 0.015\) to \(0.040\). In the regime \(d/l\ll1\) the phase is controlled by two peaks, a left peak near \(b\simeq0.55\,d\) and a right peak near \(b\simeq4\,d\), separated by the interior zero near \(b\simeq1.3\,d\). The thick curve in Fig.~\ref{fig:TSMZbOphase} traces the global maximum of the absolute phase for each \(d/l\). For small separations the right peak is taller, since its leading coefficient is the larger of the two, and the global maximum therefore lies at \(b/l\simeq4\,d/l\). As \(d/l\) increases, however, the right peak diminishes in height faster than the left peak. The two become equal near \(d/l\simeq0.025\). Above this threshold the left peak becomes the global maximum, so the optimal geometry jumps from \(b/l\simeq4\,d/l\) down to \(b/l\simeq0.55\,d/l\). This competition between the two branches is the physical origin of the abrupt jump. It confirms that the discontinuity is not a numerical artifact, but a fundamental geometric property of the TSMZ configuration, and implies that an experimental TSMZ setup would be highly sensitive to small variations in \(d\) near this threshold, making it extremely difficult to stabilize at the optimal operating point.\\
	
	\section{Unified Comparative Phase Accumulation}
	\label{sec:unified-comparison}
	
	To synthesize the geometric optimization performed in the previous sections, we now evaluate the absolute accumulated phase \(|\Delta\phi|\) in radians for all five configurations under identical realistic parameters. It is important to note that while Figs.~\ref{fig:mz-phase-0p2} and~\ref{fig:mz-phase-0p01} plot the dimensionless rescaled phase \(|\Delta\phi|/\kappa\), Table~\ref{tab:phase_comparison} lists the physical phase, related by the prefactor, \(\kappa = Gm_1m_2/(\hbar v)\). Throughout this analysis, we assume two equal masses \(m_1=m_2=10^{-14}\,\text{kg}\). This mass scale is widely adopted in near-term proposals for testing gravitationally induced entanglement, representing a challenging but feasible target for maintaining quantum coherence over the required interaction times \cite{bose2017spin,marletto2017gravitationally,schut2023relaxation}. The interaction time is fixed at \(T=1\,\text{s}\), consistent with the coherence times envisioned for such mesoscopic superpositions. Because the total path length is given by \(l = vT\), fixing \(T\) implies that each row in Table~\ref{tab:phase_comparison} corresponds to a specific particle velocity \(v \in \{10^{-4},10^{-3}, 10^{-2}, 10^{-1}, 1\}\,\text{m/s}\), with the corresponding length \(l\) listed in the first column.
	
	The size of the spatial superposition, controlled by the vertical displacement \(b\) (for Mach--Zehnder) or \(\Delta x\) (for Stern--Gerlach), is a critical experimental parameter. Recent proposals suggest superpositions on the order of \(1\) to \(10\,\mu\text{m}\) \cite{bose2026white}. Therefore, we evaluate the phase at two representative fixed displacements, \(1\,\mu\text{m}\) and \(10\,\mu\text{m}\). Previous studies have established that keeping the Casimir-Polder interaction an order of magnitude below the gravitational interaction requires a separation of \(d \gtrsim 200\,\mu\mathrm{m}\), for the nanodiamonds \cite{di2024bose}. It has recently been proposed that employing a shielding mechanism can reduce this required separation to \(d=30\,\mu\mathrm{m}\), thereby enhancing the gravitational coupling \cite{bose2026white}. In this section, we adopt an intermediate separation of \(d=100\,\mu\mathrm{m}\) as a practical compromise. At \(v=10^{-2}\) m/s, this corresponds to a normalized separation of \(d/l = 0.01\), matching the geometry plotted in Fig.~\ref{fig:mz-phase-0p01}. Table~\ref{tab:phase_comparison} summarizes the results, comparing the absolute phases at fixed geometries against the optimized maximum phase.
	
\begin{table}[t]
	\centering
	\caption{Absolute accumulated phase \(|\Delta\phi|\) in radians for different interferometer geometries at a separation of \(d = 100\,\mu\text{m}\), assuming two equal masses \(m_1=m_2=10^{-14}\,\text{kg}\). The columns list the phase for fixed vertical displacements of \(1\,\mu\text{m}\) and \(10\,\mu\text{m}\), denoted by \(b\) or \(\Delta x\) depending on the geometry, alongside the optimized maximum phase \(|\Delta\phi|_{\text{opt}}\) and its corresponding optimal geometric ratio. The optimal ratio is defined as \((\Delta x/l)_{\text{opt}}\) for the SG configurations and \((b/l)_{\text{opt}}\) for the MZ configurations.}
	\label{tab:phase_comparison}
	\footnotesize
	\setlength{\tabcolsep}{2pt} 
	\renewcommand{\arraystretch}{1.1} 
	\begin{tabular}{@{} l c c c c @{}}
		\hline
		\(l\) (m) & \(|\Delta\phi|\) & \(|\Delta\phi|\) & \(|\Delta\phi|_{\text{opt}}\) & Opt. Ratio \\
		{[\(v\) (m/s)]} & at \(1\,\mu\)m & at \(10\,\mu\)m & & \\
		\hline
		\multicolumn{5}{l}{\textit{I. SG Linear (\(\theta = \pi/2\)), ratio: \((\Delta x/l)_{\text{opt}}\)}} \\
		\hline
		\(10^{-4}\) & \(1.22 \times 10^{-4}\) & \(8.95 \times 10^{-3}\) & \(0.078\) & \(0.707\) \\
		\(10^{-3}\) & \(1.23 \times 10^{-4}\) & \(9.53 \times 10^{-3}\) & \(0.314\) & \(0.325\) \\
		\(10^{-2}\) & \(1.23 \times 10^{-4}\) & \(9.58 \times 10^{-3}\) & \(0.502\) & \(0.116\) \\
		\(10^{-1}\) & \(1.23 \times 10^{-4}\) & \(9.59 \times 10^{-3}\) & \(0.587\) & \(0.038\) \\
		\(1\)        & \(1.23 \times 10^{-4}\) & \(9.59 \times 10^{-3}\) & \(0.618\) & \(0.012\) \\
		\hline
		\multicolumn{5}{l}{\textit{II. SG Parallel (\(\theta = \pi/2\)), ratio: \((\Delta x/l)_{\text{opt}}\)}} \\
		\hline
		\(10^{-4}\) & \(6.29 \times 10^{-5}\) & \(5.86 \times 10^{-3}\) & \(0.150 \) & \(0.990\) \\
		\(10^{-3}\) & \(6.32 \times 10^{-5}\) & \(6.24 \times 10^{-3}\) & \(0.886\) & \(0.990\) \\
		\(10^{-2}\) & \(6.33 \times 10^{-5}\) & \(6.28 \times 10^{-3}\) & \(1.199\) & \(0.990\) \\
		\(10^{-1}\) & \(6.33 \times 10^{-5}\) & \(6.28 \times 10^{-3}\) & \(1.256\) & \(0.990\) \\
		\(1\)        & \(6.33 \times 10^{-5}\) & \(6.28 \times 10^{-3}\) & \(1.265\) & \(0.990\) \\
		\hline
		\multicolumn{5}{l}{\textit{III. MZ TSMZ, ratio: \((b/l)_{\text{opt}}\)}} \\
		\hline
		\(10^{-4}\) & \(6.06 \times 10^{-5}\) & \(4.01 \times 10^{-3}\) & \(0.0137\) & \(0.365\) \\
		\(10^{-3}\) & \(6.13 \times 10^{-5}\) & \(4.57 \times 10^{-3}\) & \(0.0277\) & \(0.054\) \\
		\(10^{-2}\) & \(6.14 \times 10^{-5}\) & \(4.62 \times 10^{-3}\) & \(0.0327\) & \(0.038\) \\
		\(10^{-1}\) & \(6.14 \times 10^{-5}\) & \(4.63 \times 10^{-3}\) & \(0.0351\) & \(0.004\) \\
		\(1\)        & \(6.14 \times 10^{-5}\) & \(4.63 \times 10^{-3}\) & \(0.0354\) & \(0.0004\) \\
		\hline
		\multicolumn{5}{l}{\textit{IV. MZ MSMZ, ratio: \((b/l)_{\text{opt}}\)}} \\
		\hline
		\(10^{-4}\) & \(1.82 \times 10^{-4}\) & \(1.24 \times 10^{-2}\) & \(0.054\) & \(0.411\) \\
		\(10^{-3}\) & \(1.84 \times 10^{-4}\) & \(1.41 \times 10^{-2}\) & \(0.311\) & \(0.232\) \\
		\(10^{-2}\) & \(1.84 \times 10^{-4}\) & \(1.43 \times 10^{-2}\) & \(0.515\) & \(0.089\) \\
		\(10^{-1}\) & \(1.84 \times 10^{-4}\) & \(1.43 \times 10^{-2}\) & \(0.595\) & \(0.030\) \\
		\(1\)        & \(1.84 \times 10^{-4}\) & \(1.43 \times 10^{-2}\) & \(0.621\) & \(0.010\) \\
		\hline
		\multicolumn{5}{l}{\textit{V. MZ PAMZ, ratio: \((b/l)_{\text{opt}}\)}} \\
		\hline
		\(10^{-4}\) & \(1.25 \times 10^{-4}\) & \(1.08 \times 10^{-2}\) & \(0.087\) & \(0.500\) \\
		\(10^{-3}\) & \(1.26 \times 10^{-4}\) & \(1.23 \times 10^{-2}\) & \(0.791\) & \(0.500\) \\
		\(10^{-2}\) & \(1.27 \times 10^{-4}\) & \(1.25 \times 10^{-2}\) & \(1.177\) & \(0.500\) \\
		\(10^{-1}\) & \(1.27 \times 10^{-4}\) & \(1.25 \times 10^{-2}\) & \(1.253\) & \(0.500\) \\
		\(1\)        & \(1.27 \times 10^{-4}\) & \(1.25 \times 10^{-2}\) & \(1.264\) & \(0.500\) \\
		\hline
	\end{tabular}
\end{table}
	
	Several critical conclusions emerge from the exact phase calculations presented in Table~\ref{tab:phase_comparison}. First, fixing the geometric parameters without optimization is highly detrimental to the signal. For instance, at \(v=10^{-2}\)~m/s, setting a fixed gap of \(1\,\mu\)m yields a phase of only \(\sim 10^{-4}\) to \(10^{-5}\) rad across all configurations. Even increasing the gap to \(10\,\mu\)m only pushes the phase to the \(10^{-2}\) rad regime. These values are orders of magnitude smaller than the optimized values. Consequently, geometric optimization of the vertical displacement is strictly required to achieve an order-unity phase.\\
	When optimized, a clear hierarchy emerges. The SG Parallel and PAMZ configurations dominate, both achieving maximum phases exceeding \(1.1\) radians. At \(v=10^{-2}\) m/s, the SG Parallel phase is \(1.199\) rad, while the PAMZ phase is \(1.177\) rad. The difference between the two is only \(1.8\%\), which is negligible compared to the orders-of-magnitude gap separating them from the remaining configurations. The SG Linear and MSMZ setups plateau at roughly half this value, around \(0.5\) to \(0.6\) rad, while the TSMZ setup performs poorly, failing to exceed \(0.04\) rad. In particular, the PAMZ phase substantially exceeds the TSMZ phase, achieving a \(36\)-fold enhancement at this velocity.\\
	Another profound distinction lies in the stability of the optimal geometry. For the SG Linear and MSMZ setups, the optimal ratio shrinks rapidly as the path length \(l\) increases. Even at a moderate velocity of \(v=1\) m/s, the optimal ratio drops to \(0.01\), necessitating a vertical arm \(100\) times smaller than the horizontal arm. Such an extreme aspect ratio poses significant experimental challenges. In contrast, the SG Parallel setup achieves its maximum phase in the limit \((\Delta x/l)_{\text{opt}} \to 1\), which corresponds to the vanishing of the horizontal arm (\(a \to 0\)). However, as noted in the previous section, its phase reaches the plateau region much more rapidly. Consequently, even at the optimal operating point of the MSMZ setup, the SG Parallel configuration exhibits a higher accumulated phase and allows for much greater flexibility in geometric tuning. \\
	The PAMZ geometry emerges as one of the two unequivocally preferred configurations. Its optimal geometry is permanently fixed at \((b/l)_{\text{opt}} = 0.5\) across the entire parameter space. Because \(b/l=0.5\) represents a perfectly symmetric 1:1 aspect ratio, the PAMZ setup guarantees maximal phase accumulation without requiring any real-time geometric adjustments as the particle velocity or total path length is varied. This geometric stability substantially relaxes the fabrication tolerances required for the initial construction of the interferometer.  Conversely, the TSMZ configuration is the least favorable within the MZ class. It generates a weak phase, and its optimal operating points typically reside near a null region, which renders the interferometer highly unstable. \\
	Beyond theoretical phase maximization, the practical choice between these configurations will be heavily dictated by the experimentally achievable limits. Since the total path length is constrained by \(l = vT\), the maximum attainable vertical length directly dictates the required particle velocity. If future advancements permit larger vertical displacements and longer coherence times, experiments could operate at higher particle velocities, such as those exceeding \(10^{-3}\ \mathrm{m/s}\). Conversely, until such technology matures, experiments may need to employ lower velocities and smaller total path lengths to achieve the optimal geometric ratios while fitting within a laboratory-scale apparatus. Consequently, the ultimate choice of configuration will be a dynamic balance between the theoretical phase limits identified in this work and the practical constraints of current and near-future interferometric technology.
	
	\section{Impact of Geometric Tolerances on Phase Accumulation}
	\label{sec:robustness}
	
In practical implementations involving quantum particles, achieving perfect control over the system is inherently challenging. Real-world setups inevitably suffer from fabrication imperfections and environmental fluctuations, which introduce unavoidable uncertainties into the physical dimensions of each interferometer path. Since the accumulated phase depends sensitively on these geometric parameters, it is crucial to analyze how such uncertainties propagate into the final measurement. In this section, we evaluate the robustness of the accumulated phase against these practical limitations.

To model these imperfections, we assign independent \(5\%\) relative uncertainties to the horizontal arm length \(a\), the vertical displacement, and the interferometer separation \(d\) via a  Monte Carlo approach. The total path length \(l\) is treated as a derived quantity that fluctuates as a consequence of these independent geometric variations. This percentage-based approach is physically motivated because a particle traversing a longer path accumulates proportionally more absolute error due to thermal fluctuations and deviations from an ideal straight geometry. We quantify the robustness of each configuration using the relative phase uncertainty, defined as \(\delta = \sigma_{\Delta\phi}/\Delta\phi\), which allows us to systematically evaluate geometric stability across the relevant parameter space.
	\subsection{Stern--Gerlach Configurations}
	\label{sec:robustness-sg}
	
\begin{figure}[t]
		\centering
		\includegraphics[width=0.49\textwidth]{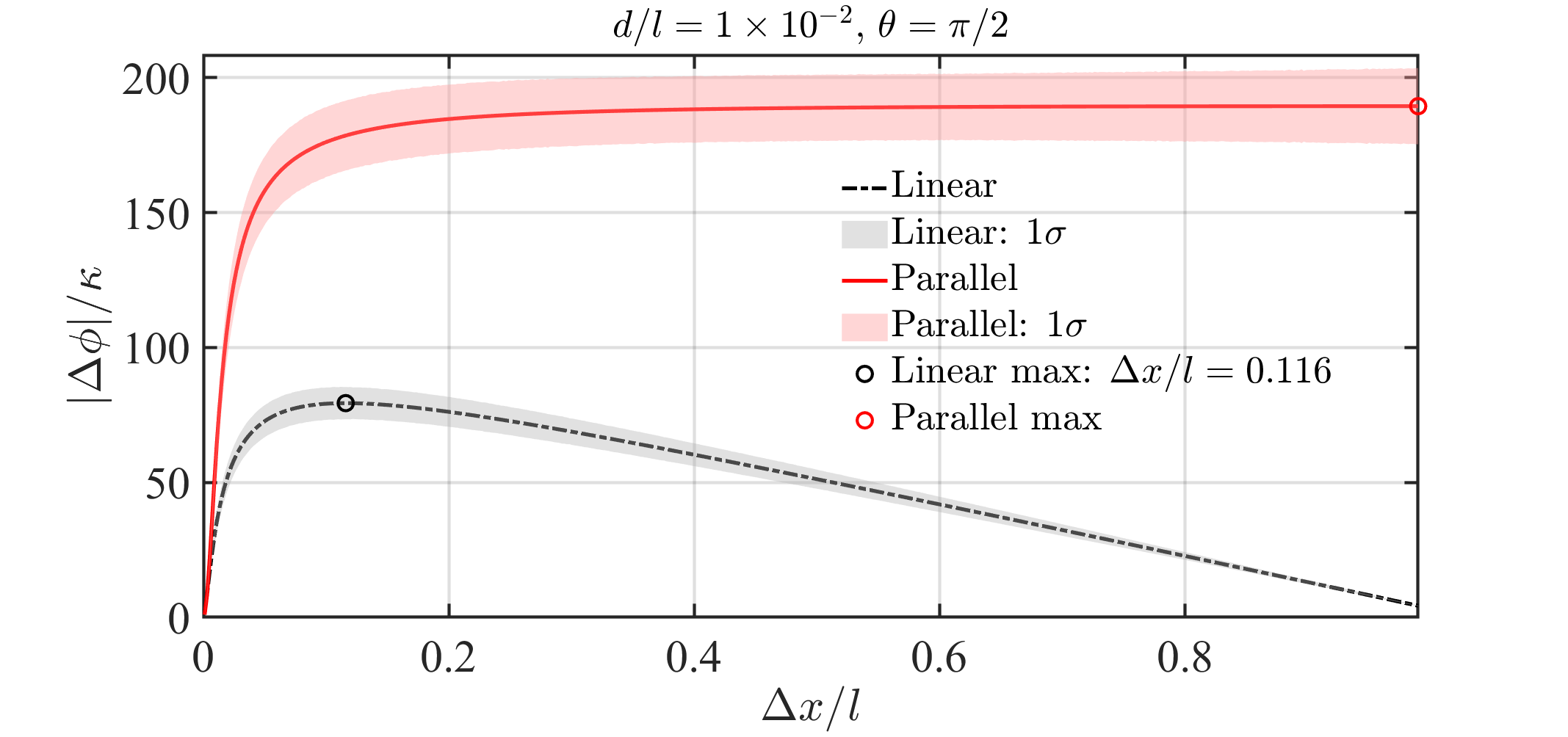}
		\caption{Phase \(|\Delta\phi|/\kappa\) and \(1\sigma\) uncertainty bands versus \(\Delta x/l\) for the SG configurations at \(d/l=0.01\) and \(\theta=\pi/2\). The solid curves represent the parallel configuration, and the dash-dotted curves represent the linear configuration. Shaded regions indicate the \(1\sigma\) uncertainty arising from \(5\%\) variations in \(\Delta x\), \(a\)  and \(d\). The parallel band grows monotonically, while the linear band exhibits a maximum band width at its phase peak.}
		\label{fig:sg-err-dx}
	\end{figure}

\begin{figure}[t]
		\centering
		\includegraphics[width=0.49\textwidth]{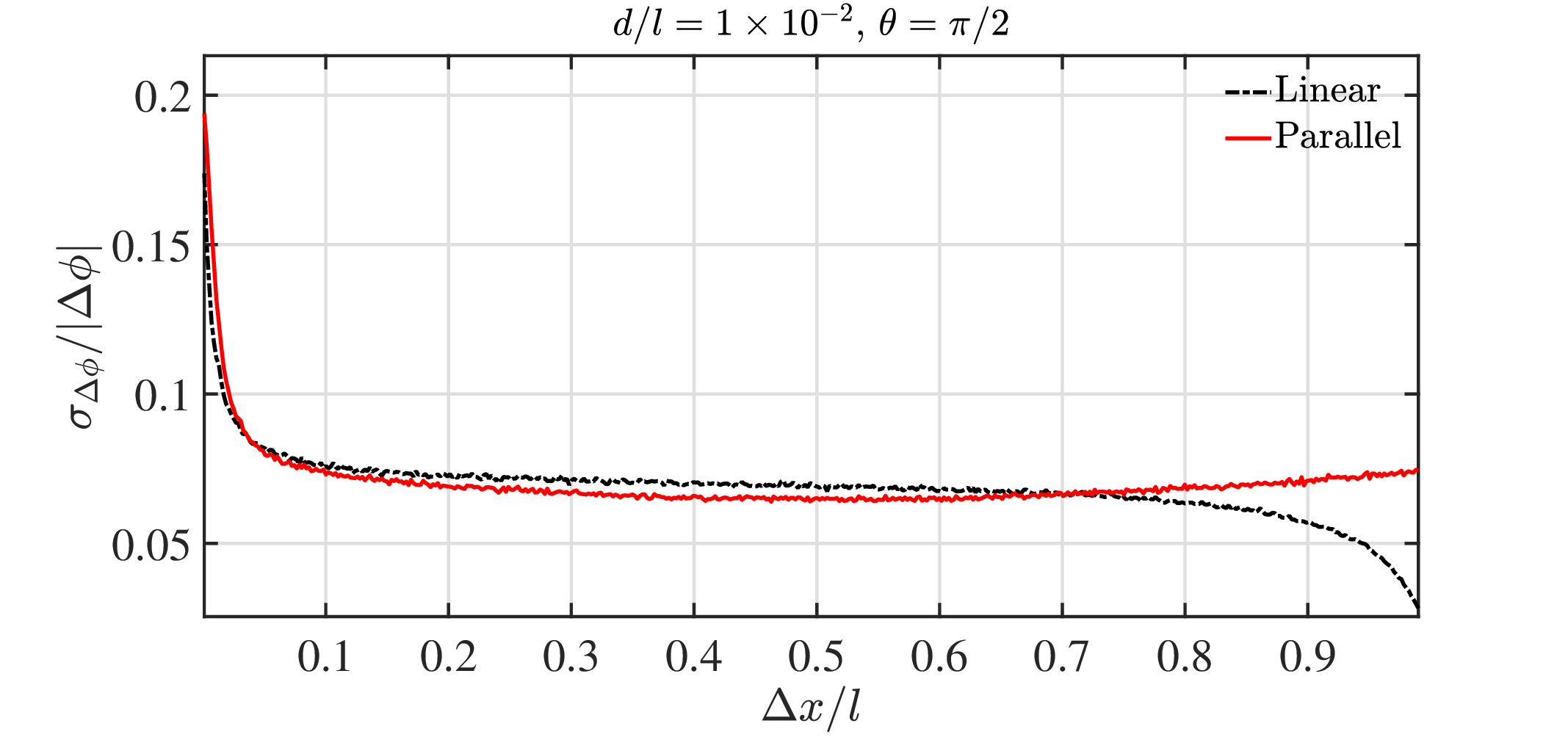}
		\caption{Relative uncertainty \(\sigma_{\Delta\phi}/|\Delta\phi|\) versus \(\Delta x/l\) for the SG configurations at \(d/l=0.01\) and \(\theta=\pi/2\). The linear configuration (dash-dotted) exhibits a nearly constant behavior in the intermediate \(\Delta x/l\) range before decreasing sharply at larger values. The parallel configuration (solid) maintains a consistently low relative uncertainty across a broad plateau region.}
		\label{fig:sg-rel-dx}
	\end{figure}

	Figure~\ref{fig:sg-err-dx} shows the phase \(|\Delta\phi|/\kappa\) and its \(1\sigma\) uncertainty band as a function of \(\Delta x/l\) for \(d/l = 0.01\) and \(\theta = \pi/2\). In the linear configuration (dash-dotted curve), the uncertainty band is narrow at small \(\Delta x/l\), widens near the phase maximum at \(\Delta x/l \simeq 0.12\), and narrows  again at larger \(\Delta x/l\) where the phase decays. In the parallel configuration (solid curve), the absolute uncertainty grows monotonically with \(\Delta x/l\) across the entire range. Consequently, at the optimal operating point of each configuration, the parallel band is broader in absolute terms than the linear one, despite the larger accumulated phase. 
The corresponding relative uncertainties are shown in Fig.~\ref{fig:sg-rel-dx}. Both configurations exhibit large values as \(\Delta x/l \to 0\). As \(\Delta x/l\) increases, the relative uncertainty for both setups drops rapidly. At small \(\Delta x/l\), both configurations exhibit comparable relative uncertainties. The linear configuration reaches  approximately \(0.075\) near its phase optimum at \(\Delta x/l \simeq 0.12\). The parallel configuration settles into a broad, stable plateau between approximately \(0.065\) and \(0.075\) across most of the parameter space. The two curves cross near \(\Delta x/l = 0.71\). Below this value, the parallel configuration generally maintains a lower relative uncertainty. Above this value, the linear configuration exhibits a sharp drop as \(\Delta x/l \to 1\).

Combining Figs.~\ref{fig:sg-err-dx} and~\ref{fig:sg-rel-dx}, we conclude that both configurations offer robust operating regimes. The parallel configuration offers a broad plateau of low relative uncertainty that coincides with its maximum phase region. The linear configuration performs almost identically to the parallel setup near its phase optimum. However, at the extreme limit where the horizontal arm vanishes, the relative uncertainty of the linear setup becomes even smaller, even though its phase also decreases.
	
\begin{figure}[t]
		\centering
		\includegraphics[width=0.49\textwidth]{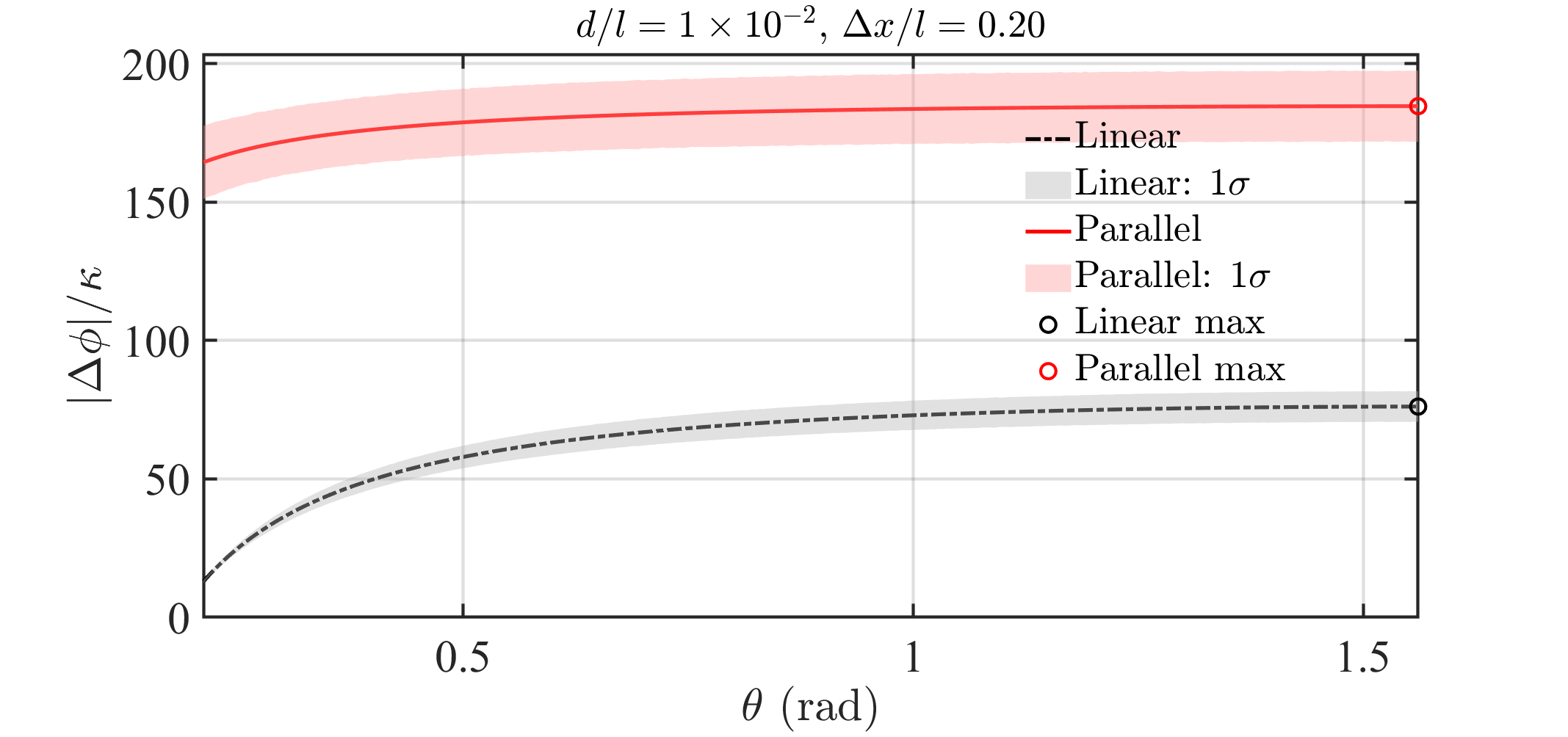}
		\caption{Phase \(|\Delta\phi|/\kappa\) and \(1\sigma\) uncertainty bands versus \(\theta\) for the SG configurations at \(d/l=0.01\) and \(\Delta x/l=0.20\). The solid curves represent the parallel configuration, and the dash-dotted curves represent the linear configuration. The uncertainty band for the linear configuration grows rapidly as \(\theta\) approaches \(\pi/2\), while the parallel band remains relatively stable.}
		\label{fig:sg-err-theta}
	\end{figure}

	Figure~\ref{fig:sg-err-theta} shows the same comparison as a function of the deviation angle \(\theta\), at fixed \(\Delta x/l = 0.20\). As \(\theta\) increases from its lower bound \(\theta_{\min} = \arcsin(\Delta x/l) \simeq 0.20\)~rad toward \(\pi/2\), the uncertainty band of the linear configuration grows. The parallel configuration exhibits the same monotonically increasing trend in band width with \(\theta\), but its band is broader at all angles, reaching more than twice the width of the linear configuration at \(\theta = \pi/2\). Notably, the band width in the parallel configuration grows more gradually, varying only slightly over the entire accessible range of \(\theta\). 
	
\begin{figure}[t]
		\centering
		\includegraphics[width=0.49\textwidth]{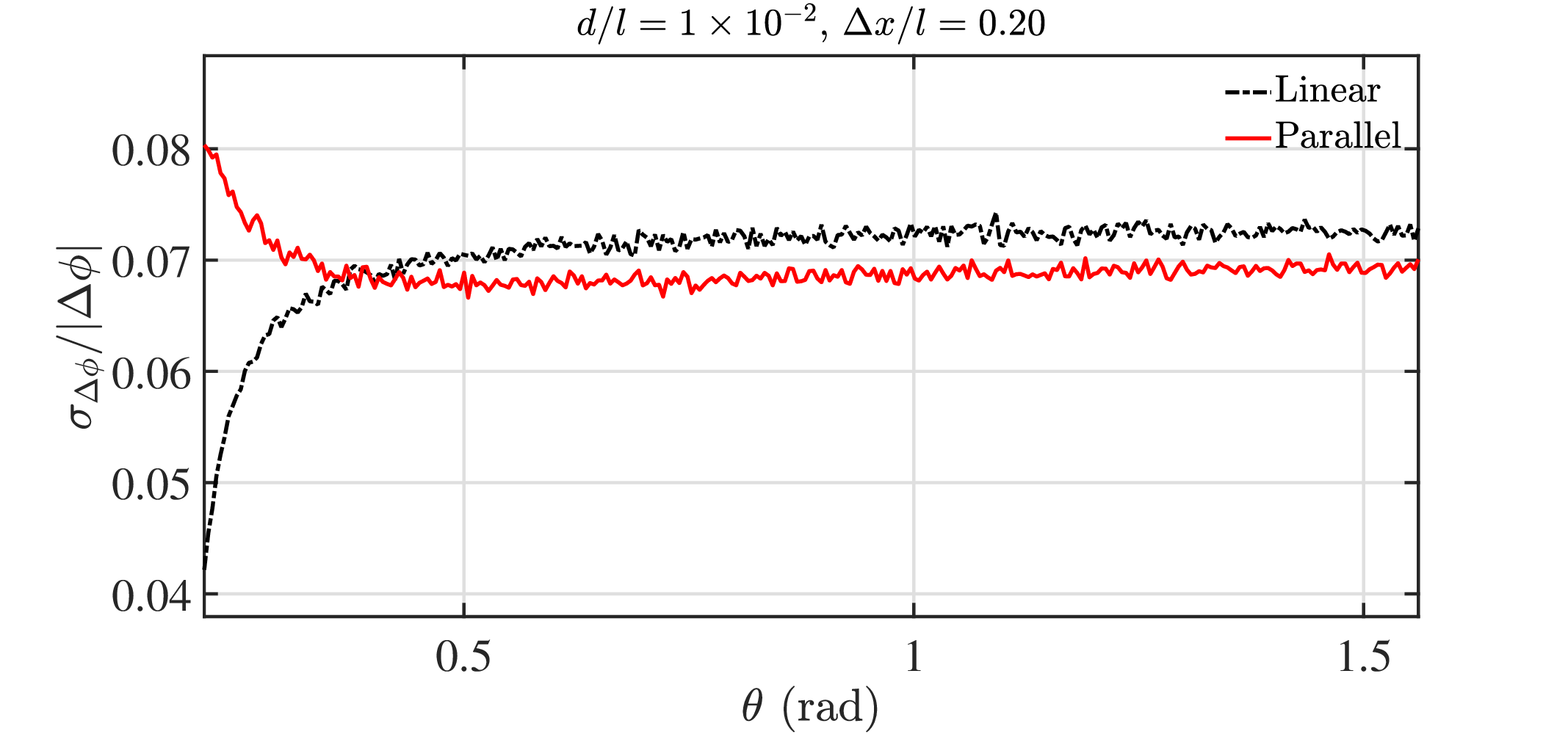}
	\caption{Relative uncertainty \(\sigma_{\Delta\phi}/|\Delta\phi|\) versus \(\theta\) for the SG configurations at \(d/l=0.01\) and \(\Delta x/l=0.20\). The linear configuration (dash-dotted) shows a steep increase in relative uncertainty as \(\theta\) decreases toward its geometric lower bound. In contrast, the parallel configuration (solid) decreases slightly from the lower bound and then remains nearly constant, with a slightly lower relative uncertainty at larger values of \(\theta\).}
		\label{fig:sg-rel-theta}
	\end{figure}

The corresponding relative uncertainties are shown in Fig.~\ref{fig:sg-rel-theta}. At the geometric lower bound \(\theta_{\min} = \arcsin(\Delta x/l) \simeq 0.21\) rad, the linear configuration exhibits a relatively low uncertainty of approximately \(0.04\), while the parallel configuration starts at a higher value of roughly \(0.08\). As \(\theta\) increases, the two configurations exhibit opposite trends. The linear configuration's relative uncertainty rises steeply and then plateaus near \(0.073\) as \(\theta \to \pi/2\). In contrast, the parallel configuration's relative uncertainty decreases and settles into a stable plateau of approximately \(0.068\) over the majority of the accessible range. The two curves cross near \(\theta \simeq 0.38\) rad. At the optimal operating angle of \(\theta = \pi/2\), both configurations provide stable relative uncertainties, with the difference between them being less than \(10\%\).

The error analysis therefore leads to a clear conclusion. Near the optimal operating angle of \(\theta = \pi/2\), the parallel configuration offers a larger accumulated phase while maintaining a stable relative uncertainty that is slightly lower than that of the linear setup. Although the linear configuration exhibits a lower accumulated phase, it achieves a comparable relative uncertainty at its optimal operating point.
	
\subsection{Mach--Zehnder Configurations}
\label{sec:robustness-mz}
\begin{figure}[t]
	\centering
	\includegraphics[width=\columnwidth]{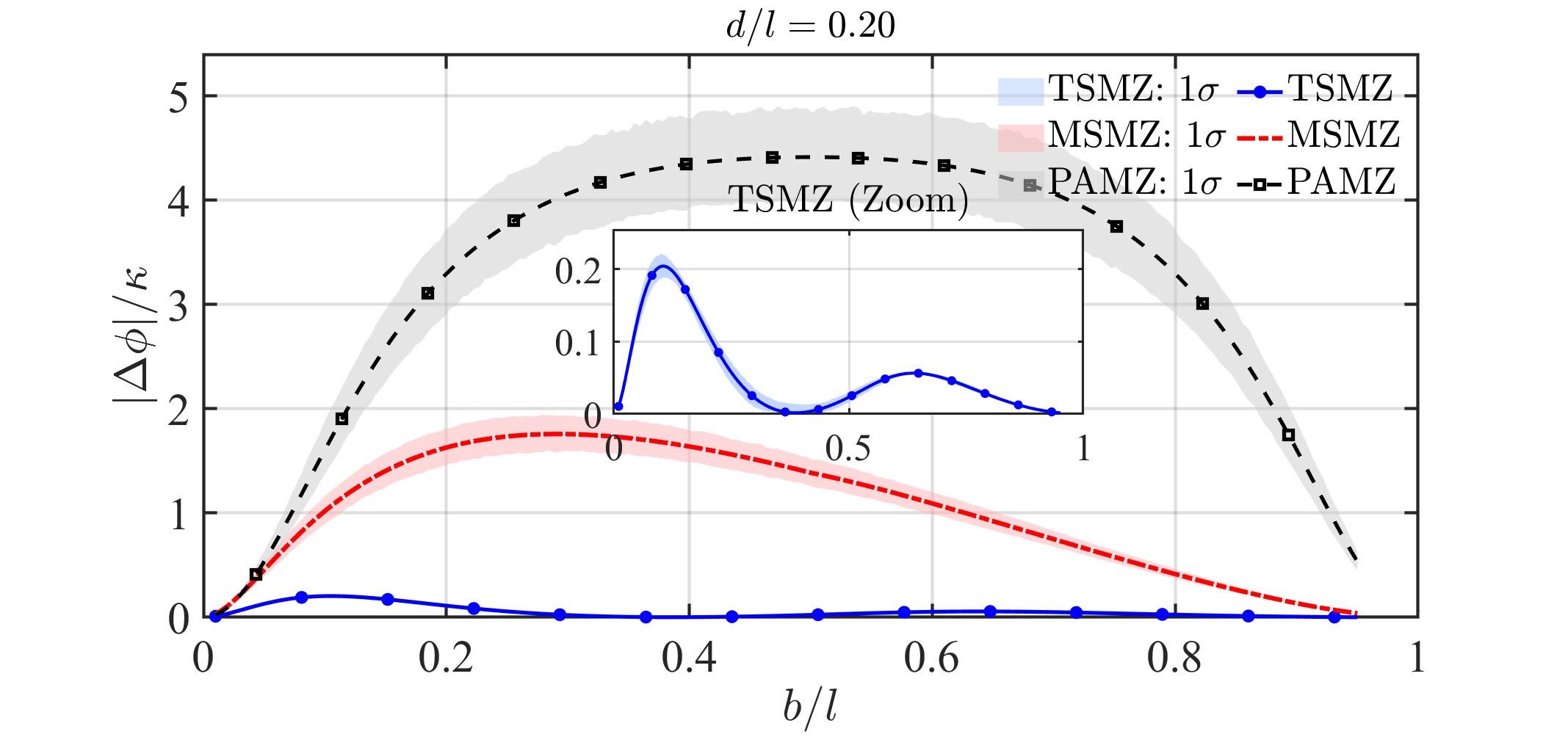}
	\caption{Phase \(|\Delta\phi|/\kappa\) and \(1\sigma\) uncertainty bands versus \(b/l\) for the MZ configurations at \(d/l = 0.20\). The TSMZ setup is shown with circle markers, the MSMZ setup with a dash-dotted curve, and the PAMZ setup with square markers. Shaded regions indicate the uncertainty from \(5\%\) variations in \(a\), \(b\) and \(d\). The inset magnifies the TSMZ curve to reveal that its absolute band remains finite at the phase nulls.}
	\label{fig:MZ0p1ERR}
\end{figure}

\begin{figure}[t]
	\centering
	\includegraphics[width=\columnwidth]{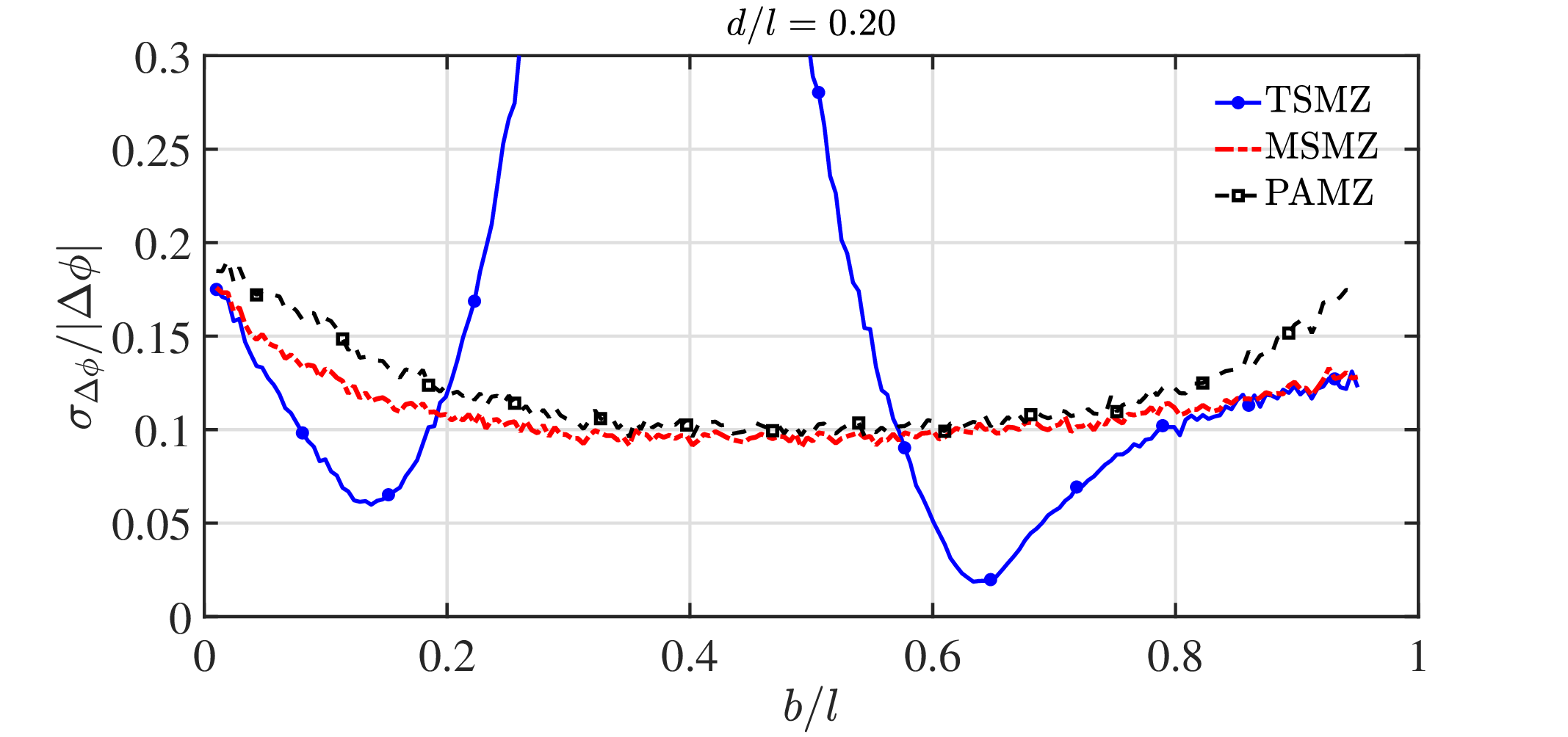}
\caption{Relative uncertainty \(\sigma_{\Delta\phi}/|\Delta\phi|\) versus \(b/l\) for the three MZ configurations at \(d/l = 0.20\). Line styles and markers are identical to Fig.~\ref{fig:MZ0p1ERR}. The TSMZ nulls cause severe uncertainty spikes, while the PAMZ and MSMZ configurations exhibit a comparable, nearly constant relative uncertainty across most of the parameter space.}
	\label{fig:MZ0p1ERReative}
\end{figure}

\begin{figure}[t]
	\centering
	\includegraphics[width=\columnwidth]{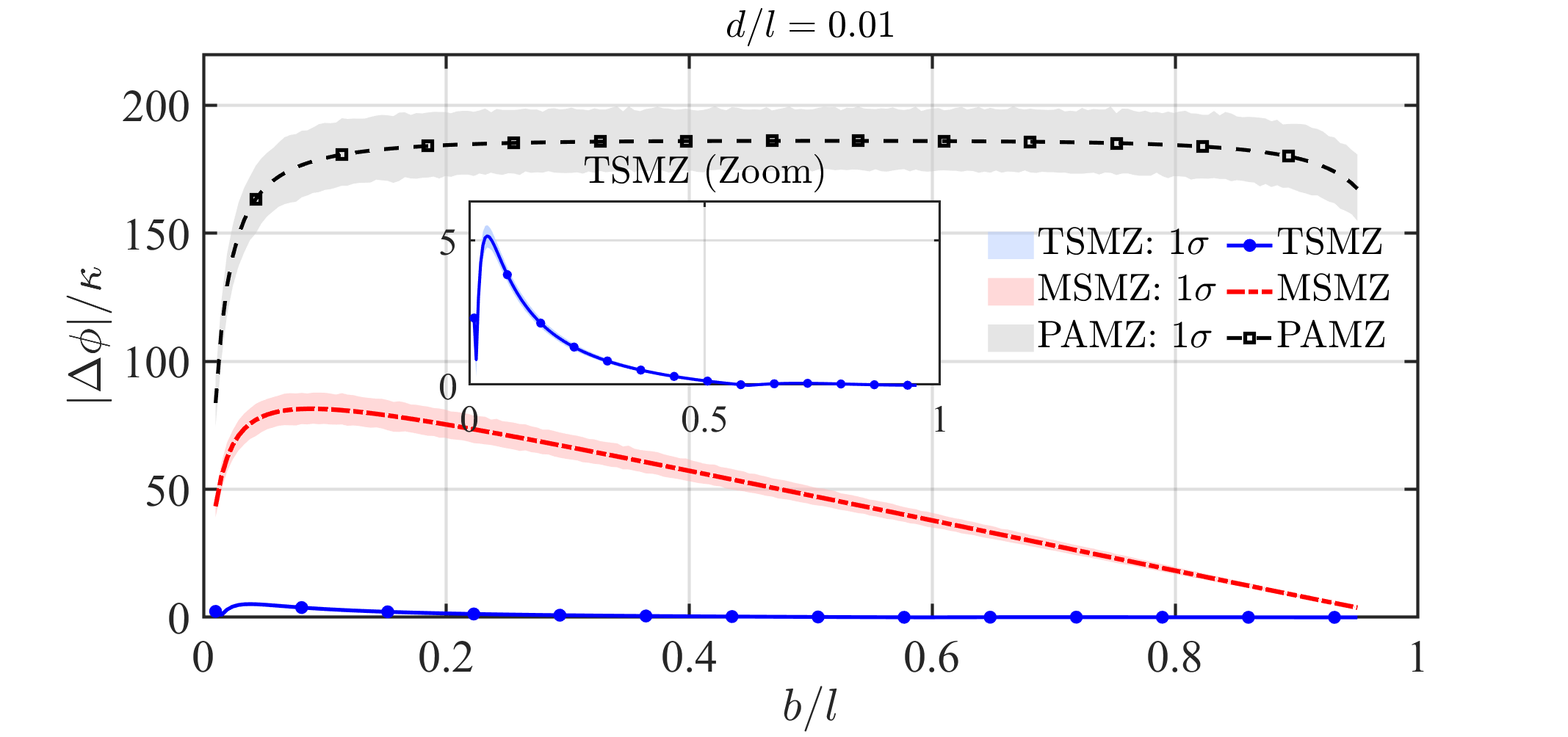}
	\caption{Phase \(|\Delta\phi|/\kappa\) and \(1\sigma\) uncertainty bands versus \(b/l\) for the MZ configurations at \(d/l = 0.01\). The line and marker styles are identical to Fig.~\ref{fig:MZ0p1ERR}. The absolute uncertainty band for PAMZ remains constant across its broad maximal phase operating window.}
	\label{fig:MZ0p01ERR}
\end{figure}

\begin{figure}[t]
	\centering
	\includegraphics[width=\columnwidth]{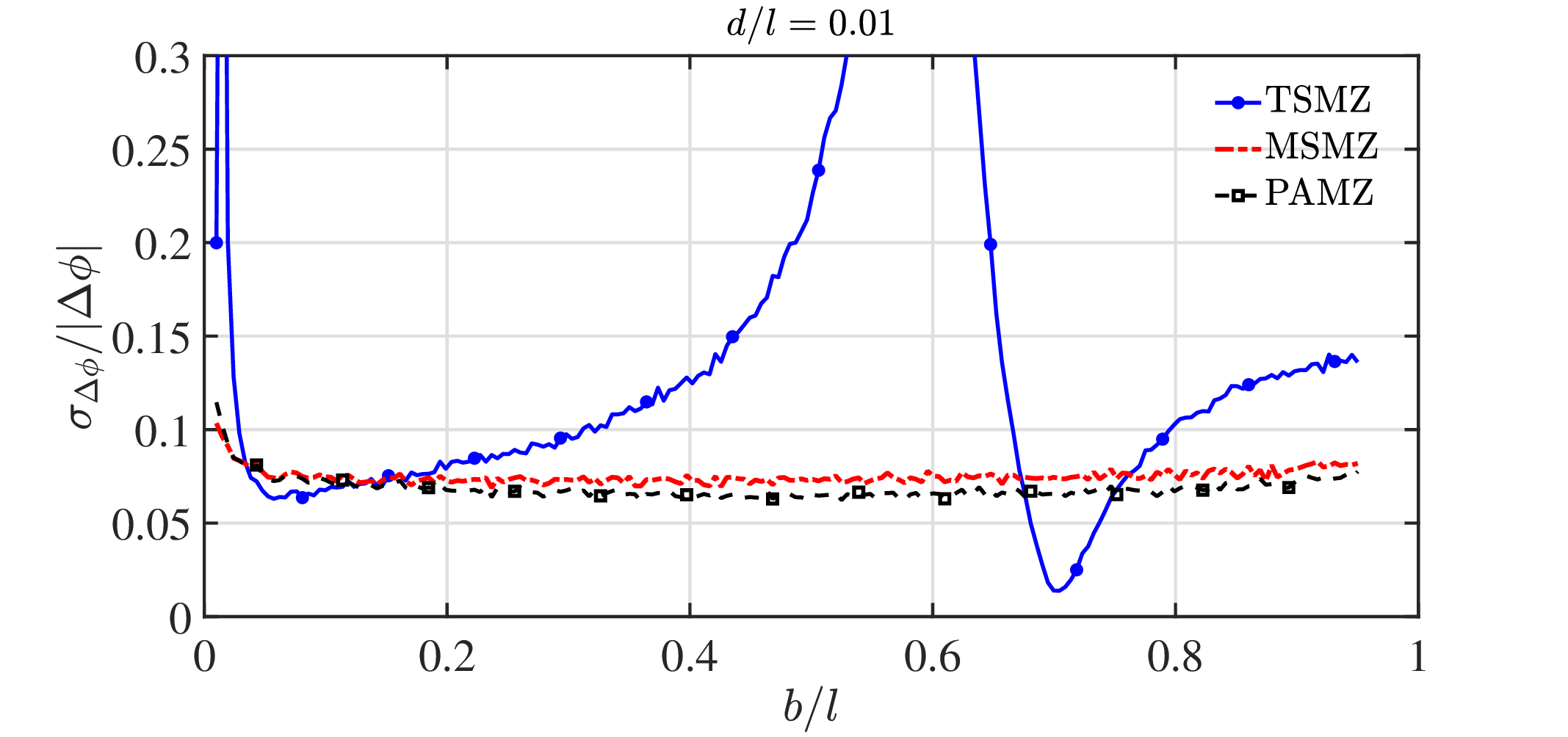}
	\caption{Relative uncertainty \(\sigma_{\Delta\phi}/|\Delta\phi|\) versus \(b/l\) for the three MZ configurations at \(d/l = 0.01\). The line styles and markers are identical to Fig.~\ref{fig:MZ0p1ERR}. The PAMZ and MSMZ configurations exhibit comparably low relative uncertainties over the broad plateau region \(0.2 \lesssim b/l \lesssim 0.8\). The TSMZ configuration diverges at its nulls but drops to the lowest uncertainty exactly at its peaks.}
	\label{fig:MZ0p01ERReative}
\end{figure}

We now perform the same uncertainty analysis for the MZ configurations. Figure~\ref{fig:MZ0p1ERR} shows the phase and its \(1\sigma\) absolute uncertainty band as a function of \(b/l\) for \(d/l = 0.20\). The three geometries display markedly different error behaviors. For MSMZ (dash-dotted curve), the uncertainty band closely tracks the phase profile and reaches its maximum near the phase peak at \(b/l \approx 0.25\). For PAMZ (dashed curve with squares), the absolute error band increases alongside the phase across the broad plateau region \(0.4 \lesssim b/l \lesssim 0.6\), where the phase reaches a peak of approximately \(4.5\). The width of the band remains relatively stable across this plateau and then gradually decreases as the phase decreases. This behavior introduces a trade-off between the magnitude of the accumulated phase and the absolute error induced by geometric imperfections. Consequently, minimizing these geometric uncertainties is particularly important for fully exploiting the advantages of the PAMZ configuration. 

The TSMZ configuration (circle markers) behaves quite differently. Its response to geometric perturbations is highly non-uniform. The absolute uncertainty is high even at the zero points where the phase reaches zero. Furthermore, its secondary phase peak at \(b/l \approx 0.65\) exhibits a noticeably narrower uncertainty band than its primary peak. The inset of Fig.~\ref{fig:MZ0p1ERR} magnifies the TSMZ curve to clearly resolve this non-trivial band structure.

The corresponding relative uncertainties are shown in Fig.~\ref{fig:MZ0p1ERReative}. Because the TSMZ absolute uncertainty band remains finite as the phase approaches zero, its relative uncertainty diverges at the phase nulls. Consequently, this behavior renders the setup highly impractical across large regions of the parameter space. Only in the immediate vicinity of the peak positions does its relative uncertainty become significantly smaller than that of the other configurations. For instance, it drops to approximately \(0.06\) near \(b/l \approx 0.14\) and to nearly \(0.02\) near \(b/l \approx 0.64\). The MSMZ relative uncertainty decreases near its phase peak and remains relatively constant, fluctuating between approximately \(0.09\) and \(0.13\) across the central region. Crucially, the PAMZ configuration maintains a relative uncertainty of approximately \(0.10\) across its central plateau \(0.4 \lesssim b/l \lesssim 0.6\), which is almost equal to that of the MSMZ setup. However, its relative uncertainty rises noticeably toward the edges of the parameter space, reaching approximately \(0.18\) as \(b/l \to 0\).

At tighter separations (\(d/l = 0.01\)), the behavior changes significantly, as shown in Figs.~\ref{fig:MZ0p01ERR} and~\ref{fig:MZ0p01ERReative}. In Fig.~\ref{fig:MZ0p01ERR}, the absolute phase for PAMZ reaches a large value of nearly \(190\) and its absolute uncertainty band remains remarkably stable and flat over a wide range of \(b/l\). Figure~\ref{fig:MZ0p01ERReative} reveals a highly competitive landscape for the relative uncertainties. Both the PAMZ and MSMZ configurations exhibit exceptionally constant and stable relative uncertainties. Although they are nearly identical in performance and both maintain a relative uncertainty between approximately \(0.067\) and \(0.078\) across the broad plateau, the PAMZ configuration achieves a slightly lower uncertainty than the MSMZ setup. In contrast, the TSMZ configuration exhibits severe divergences at its phase nulls, with the relative uncertainty spiking off the scale. While the TSMZ relative uncertainty drops to a remarkably low value of \(0.018\) exactly at its secondary peak near \(b/l \approx 0.7\), this operating window is extremely narrow. Therefore, at this tight separation, both parallel and mirror-symmetric setups offer highly stable operating regimes, with PAMZ providing a marginally lower relative uncertainty. \\
The large accumulated phase and robustness results carry direct implications for experimental design. Because the PAMZ phase is maximized over a broad range of \(b/l\), the experimentalist does not need to maintain the vertical arm separation to sub-micron precision. A \(5\%\) variation in \(d\) or thermal drift in \(b\) will not significantly degrade the accumulated phase. In contrast, the narrow phase peak of the linear SG and the sharp nulls of the TSMZ require precise geometric tuning. If thermal fluctuations cause the beam splitters to drift, a TSMZ setup could inadvertently approach a null, suppressing the signal. Thus, the PAMZ plateau acts as a geometric buffer against environmental noise, significantly relaxing fabrication tolerances.

\begin{table}[t]
	\centering
	\footnotesize
	\caption{Qualitative summary of the three Mach--Zehnder geometries at tight separations. Both the parallel-arms (PAMZ) and mirror-symmetric (MSMZ) configurations combine a large phase magnitude with a stable operating plateau.}
	\label{tab:mz-summary}
	\setlength{\tabcolsep}{4pt}
	\begin{tabular*}{\columnwidth}{@{\extracolsep{\fill}} l c c c @{}}
		\hline\hline
		Feature & TSMZ & MSMZ & PAMZ \\
		\hline
		Nulls in phase & Severe & None & None \\
		Stable optimal geometry & low & medium & high \\
		Max \(|\Delta\phi|/\kappa\) (\(d/l=0.01\)) & \(\sim 5\) & \(\sim 80\) & \(\sim 190\) \\
		Relative uncertainty & Diverges & low & low \\
		\hline\hline
	\end{tabular*}
\end{table}

Beyond geometric stability, the accumulated phase bears directly on the coherence requirements. Because PAMZ accumulates a larger phase per unit interaction time than MSMZ (by a factor of roughly \(2\) at \(d/l=0.01\)), the interaction time \(T\) required to reach a given phase is correspondingly shorter. Since the coherence time of the platform must exceed \(T\), this relaxation of \(T\) reduces the coherence-time demand, a practical advantage for the experimental realization of gravitationally induced entanglement. A qualitative summary of these MZ geometries is presented in Table~\ref{tab:mz-summary}. Ultimately, the feasibility of these proposals will depend heavily on future advancements in interferometric device precision and control.

\section{BELL VIOLATION AND OBSERVABILITY}
\label{sec:bell}

\begin{table*}[t]
	\centering
	\caption{Bell violation parameters for all five interferometer geometries at \(m = 10^{-14}\) kg, \(l = 1\) cm, \(d = 100\ \mu\)m, \(\theta = \pi/2\), and \(T = 1\) s. The optimal geometric ratio is defined as \(x_{\text{opt}} \equiv (\Delta x/l)_{\text{opt}}\) for the SG configurations and \(x_{\text{opt}} \equiv (b/l)_{\text{opt}}\) for the MZ configurations. For each configuration, the table lists the optimal geometric ratio, the rescaled phase \(|\Delta\phi|/\kappa\), the absolute phase \(\Delta\phi\), the concurrence \(\mathcal{C}\), the Bell violation \(B_{\text{max}}-2\), the relative uncertainty \(\delta\), the signal-to-noise ratio per measurement, and the minimum number of trials \(N_{5\sigma}\) required for a \(5\sigma\) detection.}
	\label{tab:bell_100um}
	\small
	\setlength{\tabcolsep}{4pt}
	\begin{tabular}{@{} l c c c c c c c c @{}}
		\hline\hline
		Config & \(x_{\text{opt}}\) & \(|\Delta\phi|/\kappa\) & \(\Delta\phi\) (rad) & \(\mathcal{C}\) & \(B_{\text{max}}-2\) & \(\delta\) & \(\text{SNR}_B\) & \(N_{5\sigma}\) \\
		\hline
		SG Linear   & \(0.120\) & \(79.4\) & \(0.502\) & \(0.248\) & \(6.08 \times 10^{-2}\) & \(0.074\) & \(6.92\) & \(0.52\) \\
		SG Parallel & \(0.800\) & \(189.3\) & \(1.198\) & \(0.564\) & \(2.96 \times 10^{-1}\) & \(0.068\) & \(8.91\) & \(0.31\) \\
		MZ TSMZ     & \(0.039\) & \(5.2\) & \(0.033\) & \(0.016\) & \(2.67 \times 10^{-4}\) & \(0.073\) & \(6.78\) & \(0.54\) \\
		MZ MSMZ     & \(0.089\) & \(81.4\) & \(0.515\) & \(0.255\) & \(6.39 \times 10^{-2}\) & \(0.075\) & \(6.90\) & \(0.53\) \\
		MZ PAMZ     & \(0.499\) & \(186.0\) & \(1.177\) & \(0.555\) & \(2.88 \times 10^{-1}\) & \(0.065\) & \(9.33\) & \(0.29\) \\
		\hline\hline
	\end{tabular}
\end{table*}

\begin{table*}[t]
	\centering
	\caption{Bell violation parameters for all five interferometer geometries at \(m = 10^{-15}\) kg, \(l = 1\) cm, \(d = 30\ \mu\)m, \(\theta = \pi/2\), and \(T = 1\) s. The column definitions are identical to Table~\ref{tab:bell_100um}.}
	\label{tab:bell_30um}
	\small
	\setlength{\tabcolsep}{4pt}
	\begin{tabular}{@{} l c c c c c c c c @{}}
		\hline\hline
		Config & \(x_{\text{opt}}\) & \(|\Delta\phi|/\kappa\) & \(\Delta\phi\) (rad) & \(\mathcal{C}\) & \(B_{\text{max}}-2\) & \(\delta\) & \(\text{SNR}_B\) & \(N_{5\sigma}\) \\
		\hline
		SG Linear   & \(0.065\) & \(293.1\) & \(0.0185\) & \(0.009\) & \(8.601 \times 10^{-5}\) & \(0.074\) & \(6.72\) & \(0.55\) \\
		SG Parallel & \(0.800\) & \(653.6\) & \(0.0414\) & \(0.021\) & \(4.28 \times 10^{-4}\) & \(0.066\) & \(7.54\) & \(0.44\) \\
		MZ TSMZ     & \(0.011\) & \(18.1\) & \(0.0011\) & \(0.001\) & \(3.29 \times 10^{-7}\) & \(0.076\) & \(6.57\) & \(0.58\) \\
		MZ MSMZ     & \(0.051\) & \(298.7\) & \(0.0189\) & \(0.009\) & \(8.94 \times 10^{-5}\) & \(0.072\) & \(6.91\) & \(0.52\) \\
		MZ PAMZ     & \(0.499\) & \(649.3\) & \(0.0411\) & \(0.021\) & \(4.22 \times 10^{-4}\) & \(0.063\) & \(7.87\) & \(0.40\) \\
		\hline\hline
	\end{tabular}
\end{table*}

To assess the experimental observability of the gravitationally induced entanglement, we evaluate the Bell parameter of Eq.~\eqref{eq:bell-param}, using \(B_{\text{max}} - 2\) as our measure of the observable quantum signature. Throughout this section, we denote by \(\delta\) the dimensionless relative uncertainty in the phase, defined as the ratio of the standard deviation of the phase to the phase itself, \(\delta = \sigma_{\Delta\phi}/\Delta\phi\). This is the same quantity computed in Sec.~\ref{sec:robustness}. The signal-to-noise ratio for the Bell parameter is defined as \(\text{SNR}_B = (B_{\text{max}}-2)/\sigma_B\), where \(\sigma_B\) is the standard deviation of \(B_{\text{max}}\) over the Monte Carlo ensemble. The minimum number of measurements required to achieve a \(5\sigma\) detection is then given by \(N_{5\sigma} = (5/\text{SNR}_B)^2\).

We consider massive particles moving at \(v = 10^{-2}\ \mathrm{m/s}\) and a deviation angle of \(\theta = \pi/2\). For each geometry, we evaluate the Bell parameter at its respective optimal operating point, \((\Delta x/l)_{\text{opt}}\) for the SG configurations and \((b/l)_{\text{opt}}\) for the MZ configurations. Because the SG parallel configuration exhibits a broad and nearly flat phase plateau, we adopt an operating point of \((\Delta x/l)_{\text{opt}} = 0.8\) for our calculations. This choice is experimentally more practical than operating at the extreme geometric limit and has a negligible effect on the accumulated phase and the resulting Bell violation.\\
 To explore how the parameter regime affects observability, we evaluate two representative cases. The first is a standard separation of \(d=100\,\mu\)m (\(d/l = 0.01\)) with \(m = 10^{-14}\) kg. The second is a tight separation of \(d=30\,\mu\)m (\(d/l = 0.003\)), achievable through Casimir--Polder screening with a conducting plate \cite{braccini2023large}; However, we pair this tighter separation with a reduced mass of \(m = 10^{-15}\) kg, a value that is experimentally much easier to achieve.

Table~\ref{tab:bell_100um} summarizes the results for the standard case, with \(m = 10^{-14}\) kg, \(l = 1\) cm, and \(T = 1\) s. The results reveal a clear hierarchy among the five configurations. The TSMZ setup performs poorly, yielding an essentially negligible violation of \(B_{\text{max}}-2 = 2.67 \times 10^{-4}\) even at the optimal geometry. The SG Linear and MSMZ setups perform moderately well, both yielding \(B_{\text{max}}-2 \approx 6 \times 10^{-2}\) with concurrences around \(0.25\). At this separation, the best-performing configurations are the SG Parallel and PAMZ setups. The PAMZ setup attains the highest signal-to-noise ratio of \(\text{SNR}_B = 9.33\), closely followed by the SG Parallel geometry with \(\text{SNR}_B = 8.91\). Both achieve a Bell violation of \(B_{\text{max}}-2 \approx 2.9 \times 10^{-1}\), which is roughly \(4.5\) to \(4.9\) times larger than that of the moderate setups. Importantly, for all five cases the required number of measurements is \(N_{5\sigma} < 1\). Assuming a \(5\%\) fabrication accuracy, this indicates that the phase uncertainty originating from geometric imperfections is sufficiently small, and the expected significance of a single measurement exceeds \(5\sigma\). Consequently, the primary experimental bottleneck is not the statistical accumulation over many trials. Instead, it is the preservation of quantum coherence throughout the interaction time and the mitigation of other non-geometric noise sources. \\

The situation changes qualitatively when the separation is reduced to \(d = 30\ \mu\)m but the mass is simultaneously lowered to \(m = 10^{-15}\) kg (Table~\ref{tab:bell_30um}). Because the gravitational phase scales as \(\kappa \propto G m^2/(\hbar v)\), the tenfold reduction in mass suppresses the phase by two orders of magnitude. The tighter separation partially offsets this through the geometric factor, but the net effect is still a suppression of the absolute phase relative to the standard case. For example, the SG Parallel rescaled phase rises from \(189.3\) at \(d = 100\ \mu\)m to \(653.6\) at \(d = 30\ \mu\)m, a geometric gain of \(3.45\), while the mass factor contributes \(10^{-2}\). Since the concurrence is small, \(\mathcal{C} \approx \Delta\phi/2\), and the Bell violation scales as \(B_{\text{max}}-2 \approx \mathcal{C}^2\), the violation is suppressed by roughly the square of this net factor, i.e., by nearly three orders of magnitude. The resulting concurrences are all below \(0.022\), and the Bell violations \(B_{\text{max}}-2\) are at most \(4.28 \times 10^{-4}\). The SG Parallel and PAMZ configurations still yield the largest violations (\(4.28 \times 10^{-4}\) and \(4.22 \times 10^{-4}\), respectively), while the SG Linear and MSMZ setups fall to approximately \(8\) to \(9 \times 10^{-5}\), and the TSMZ setup is essentially negligible at \(3.29 \times 10^{-7}\).

Notably, the SNR values remain highly clustered across all configurations, ranging between \(6.57\) and \(7.87\). This is a direct consequence of the small-phase regime. For \(\mathcal{C} \ll 1\), the uncertainty in \(B_{\text{max}}\) is \(\sigma_B \approx 2\mathcal{C}^2\delta\), so that \(\text{SNR}_B \approx 1/(2\delta)\). The SNR therefore depends only on the relative uncertainty \(\delta\), not on the absolute phase magnitude. Since \(\delta\) varies between \(0.063\) and \(0.076\) across the five configurations, the SNR remains relatively stable, and the differences primarily reflect these variations in \(\delta\). The PAMZ configuration, with the smallest relative uncertainty (\(\delta = 0.063\)), consequently attains the highest SNR of \(7.87\), while the SG Parallel setup follows with \(\delta = 0.066\) and an SNR of \(7.54\).\\
Comparing both parameter regimes reveals a crucial design principle. While all configurations yield \(N_{5\sigma} < 1\) in both sets, the physical significance of the violation differs enormously. At \(d = 100\ \mu\)m, the SG Parallel and PAMZ violations are a substantial fraction of the maximum quantum violation, known as the Cirel'son bound of \(0.828\) \cite{cirel1980quantum}, indicating strong entanglement. At \(d = 30\ \mu\)m with \(m = 10^{-15}\) kg, the violations decrease by nearly three orders of magnitude and are only about \(0.05\%\) of the Cirel'son bound, indicating extremely weak entanglement. Furthermore, in the small-mass regime, the convergence of the SNR to \(1/(2\delta)\) means that all geometries appear statistically comparable. Although the present Monte Carlo analysis specifically isolates geometric uncertainties and fabrication tolerances, future experimental implementations will inevitably encounter additional noise sources that depend on the specific particle candidate and the interferometric apparatus. Unlike statistical noise, which scales as \(1/\sqrt{N}\) with repeated measurements, these systematic errors do not cancel over time. Therefore, maximizing the absolute phase is essential to ensure that the physical signal remains resolvable. If the absolute Bell violation is suppressed below the systematic error floor, statistical repetition alone cannot render the signal observable.
	
\section{DISCUSSION AND CONCLUSIONS}
\label{sec:summary}

We have performed a systematic geometric optimization and comparative analysis of two classes of matter-wave interferometers, Stern--Gerlach and Mach--Zehnder, for generating gravitationally induced entanglement. By deriving exact analytical expressions for the gravitational phase and performing a Monte Carlo error analysis, we have identified the geometric configurations that maximize the accumulated phase while remaining robust to fabrication tolerances.

The first key result of this work is that geometric optimization is indispensable. For commonly assumed experimental parameters, fixing the geometric ratios without optimization, as is often done in the literature, can suppress the accumulated phase and hence the detectability of entanglement up to a few orders of magnitude. This demonstrates that widely used geometric approximations can lead to qualitatively incorrect conclusions regarding experimental feasibility. A full optimization of the geometric ratios is therefore a prerequisite for any realistic assessment.

The second key result is the systematic comparison of the five configurations across the parameter ranges studied. For the SG class, the parallel-arms configuration generally produces a larger phase than the linear configuration, with an enhancement factor of \(2.4\) to \(2.7\) depending on the normalized separation \(d/l\). The Monte Carlo analysis further shows that the parallel configuration maintains a broad plateau of low relative uncertainty across the accessible range of \(\Delta x/l\). The linear configuration exhibits a comparable relative uncertainty near its phase optimum.

For the MZ class, the translation-symmetric setup is uniquely problematic. Its phase profile contains an interior zero that results in divergent points and regions with very high relative uncertainty. The mirror-symmetric setup eliminates these nulls and produces a smooth, monotonic optimal geometry. However, it is generally outperformed by the parallel-arms configuration, which achieves a larger phase and maintains a broad plateau of high phase and comparable relative uncertainty over a wide range of \(b/l\).

Our Bell violation analysis at two representative separations further corroborates these findings and establishes experimental feasibility. At a standard separation of \(100\,\mu\)m, the SG Parallel and PAMZ setups yield strong violations of \(B_{\text{max}}-2 \approx 0.29\), which are roughly five times larger than the linear setups and orders of magnitude larger than the TSMZ setup. \\
The comparison of all five geometries reveals that the SG Parallel and PAMZ configurations are the two most promising candidates, with accumulated phases that are almost equivalent across the parameter space studied. The choice between them is therefore not determined solely by their theoretical figures of merit but by their experimental implementation. The SG Parallel configuration relies on spin coherence and requires a highly elongated vertical geometry. This makes it naturally suited to particles with well-defined spin degrees of freedom but experimentally demanding at the \(10^{-14}\) kg scale. The PAMZ configuration requires no spin coherence and operates via beam splitters with a symmetric geometry featuring comparable vertical and horizontal arms, while also offering a marginal statistical advantage. However, building high-quality beam splitters for such massive particles is itself at the frontier of current technology. The final choice will ultimately depend on whether spin-coherence or beam-splitter technology matures first.

In summary, our analysis establishes that full geometric optimization is essential for the design of feasible experiments probing the quantum nature of gravity. By providing larger phase signals and robust operating plateaus, parallel-arms designs offer a practical pathway toward this goal. These optimized geometries ultimately bring the experimental realization of quantum gravity tests one step closer to reality.
	\bibliography{library}
	
\end{document}